\documentclass[floats,preprint,nofootinbib,superscriptaddress,tightenlines]{revtex4}
\usepackage{amsfonts,amsmath,amssymb,slashed}
\usepackage{bm}% bold math
\usepackage[colorlinks,anchorcolor=black,citecolor=violet,linkcolor=blue]{hyperref}
\usepackage{graphicx}% Include figure files
\usepackage{xcolor}

\begin{document}
\baselineskip=17pt \parskip=3pt

\preprint{NCTS-PH/1909}
\hspace*{\fill}

\title{Exploring leptoquark effects in hyperon and kaon decays \\ with missing energy}

\author{Jhih-Ying Su}
\affiliation{Department of Physics, National Taiwan University, Taipei 106, Taiwan}

\author{Jusak Tandean}
\affiliation{Department of Physics, National Taiwan University, Taipei 106, Taiwan}
\affiliation{Physics Division, National Center for Theoretical Sciences, Hsinchu 300, Taiwan
\bigskip}

%\date{\vspace*{3ex}}

\begin{abstract}

We entertain the possibility that scalar leptoquarks (LQs) generate consequential effects on
the strangeness-changing decays of hyperons and kaons involving missing energy carried away
by a pair of invisible fermions.
Although such processes have suppressed rates in the standard model (SM), they could get
significant enhancement in the presence of the LQs.
In order to respect the available data on the kaon modes $K\to\pi\nu\bar\nu$ and increase the rates
of the hyperon decays substantially at the same time, two different scalar LQs are needed.
If the LQs have Yukawa couplings solely to SM fermions, we find that the hyperon rates cannot
attain values within the reach of ongoing or near-future experiments because of the combined
constraints from the measurements on kaon mixing and lepton-flavor-violating processes.
However, if we include light right-handed neutrinos in the LQ interactions, their contributions
can evade the leading restrictions and translate into hyperon rates which may be big enough to be
probed by upcoming searches.
Thus, these hyperon modes could provide a new avenue for seeking sterile neutrinos.

\end{abstract}

\maketitle

\section{Introduction \label{intro}}

The flavor-changing neutral current (FCNC) decays of strange hadrons involving missing energy
are of major importance because they are known to be dominated by short-distance
physics~\cite{Buchalla:1995vs}.
Therefore these decays offer beneficial access not only to the underlying weak dynamics of
the standard model (SM) but also to manifestations of possible new physics (NP) beyond it.
Such processes arise mainly from the quark transition \,$s\to d\slashed E$\, with
the missing energy ($\slashed E$) being carried away by one or more invisible particles.
In the SM this proceeds from loop diagrams~\cite{Buchalla:1995vs} and a pair of undetected
neutrinos ($\nu\bar\nu$) act as the invisibles.
This entails that the SM branching fractions of the corresponding hadron decays are very small,
which makes them good places to look for indications of NP.
If it exists, there could be additional ingredients which modify the SM amplitudes
and/or give rise to extra channels with invisible nonstandard particles.
These changes may lead to signs that are sizable enough to be discovered by running or
forthcoming searches.

To date the most intensive quests for \,$s\to d\slashed E$\, have focused on the kaon modes
\,$K\to\pi\nu\bar\nu$\, and produced upper limits on their branching
fractions~\cite{Artamonov:2008qb,Ahn:2018mvc,CortinaGil:2018fkc,Tanabashi:2018oca}
which are fairly close to the SM expectations~\cite{Tanabashi:2018oca}.
This implies that the room for NP in \,$K\to\pi\slashed E$\, can no longer be considerable.
Nevertheless, as pointed out in Refs.\,\,\cite{Tandean:2019tkm,Li:2019cbk}, the impact of NP
could still be large through operators having exclusively parity-odd $ds$ quark bilinears.
The reason is that these operators do not participate directly in \,$K\to\pi\slashed E$,\, hence
not restricted by their data, and are instead subject to restraints from the channels with no or
two pions, namely \,$K\to\slashed E$\, and \,$K\to\pi\pi'\slashed E$,\, which at the moment have
comparatively loose empirical bounds \cite{Tanabashi:2018oca,Gninenko:2014sxa,Adler:2000ic,
E391a:2011aa}.
Future efforts to improve upon the data on the latter two sets of channels would then be
highly desirable.

The effective \,$s\to d\slashed E$\, operators governing all of the aforementioned kaon modes
affect their baryon counterparts as well and consequently may also be testable in the FCNC
decays of hyperons with missing energy \cite{Tandean:2019tkm,Li:2019cbk},
on which there has been no measurement yet~\cite{Tanabashi:2018oca}.
It is therefore of keen interest that there has recently been a proposal to pursue them in
the BESIII experiment~\cite{Li:2016tlt}, which is currently in operation and has acquired
an ample hyperon dataset~\cite{Ablikim:2019hff}.
As demonstrated in~Refs.\,\,\cite{Tandean:2019tkm,Li:2019cbk}, the constraints from
\,$K\to\slashed E$\, and \,$K\to\pi\pi'\slashed E$\, are sufficiently loose to allow these
hyperon modes to have branching fractions greatly exceeding their SM predictions and within
the proposed sensitivity reach of BESIII for these processes\,\,\cite{Li:2016tlt}.
Its search for them may then be expected to yield valuable information regarding NP which could
be lurking in the underlying quark transition.

The foregoing motivates us in this paper to examine these strangeness-changing ($|\Delta S|=1$)
hyperon decays with missing energy in the contexts of relatively simple NP models.
It also serves to complement earlier model-independent analyses in which the missing energy
was assumed to be carried away by an invisible pair of spin-1/2 particles~\cite{Tandean:2019tkm}
or spinless ones~\cite{Li:2019cbk}.
Within a\,\,specific model the parameters determining the strength of the \,$s\to d\slashed E$\,
operators often enter other observables, which may be well constrained by their respective data.
In exploring this, one could thus learn how the different restrictions could probe various
aspects of a model and what modifications to it, if still feasible, may need to be made to
comply with the constraints.
Moreover, this exercise could provide rough guidance about what kind of NP might be responsible
if certain clear signatures beyond the SM appear from the hunts for these hyperon modes.

Here we entertain in particular the possibility that heavy leptoquarks (LQs) with spin 0 mediate
the NP contributions to the FCNCs.
Over the last few years LQs have gained plenty of attention in the literature
({\it e.g.}\,\,\cite{Bhattacharya:2016mcc,Crivellin:2017zlb,Buttazzo:2017ixm,Sahoo:2015pzk,
Sahoo:2018ffv,Kumar:2016omp,Dorsner:2016wpm,Cai:2017wry,Fajfer:2018bfj,Fayyazuddin:2018zww})
because proposed models containing them are among those that could offer the preferred
explanations for the so-called $B$-physics anomalies.
Thus, while extra data are awaited to establish whether some or all of these anomalies originate
from NP, it is germane to investigate if LQs can bring about appreciable manifestations
in the hyperon sector as well.
Furthermore, such a hyperon study has never been conducted before as far as we know,
and it is timely, now that we anticipate the aforesaid hyperon measurements by BESIII.
In contrast, potential LQ effects on \,$K\to\pi\slashed E$\, and several other kaon transitions
have been extensively treated in the past~\cite{Bobeth:2017ecx,Shanker:1982nd,Shanker:1981mj,
Buchmuller:1986iq,Davies:1990sc,Leurer:1993em,Leurer:1993qx,Davidson:1993qk,Valencia:1994cj,
Saha:2010vw,Davidson:2010uu,Sahoo:2015pzk,Sahoo:2018ffv,Kumar:2016omp,Dorsner:2016wpm,
Cai:2017wry,Fajfer:2018bfj,Fayyazuddin:2018zww,He:2018uey,He:2019xxp,Mandal:2019gff}.

Besides the scalar LQs, we will include light right-handed (RH) neutrinos in the theory.
These are singlets under the SM gauge groups and, via the LQs, can have
renormalizable links to SM quarks.
Our introduction of RH neutrinos is well motivated on a couple of grounds.
First, their existence will be required if empirical research in the future proves that
neutrinos are Dirac in nature.
Second, as will be seen later on, compared to those with only SM fermions the FCNCs involving
the RH neutrinos and SM quarks might have substantially amplified influence on the hyperon decays.

The organization of the rest of this article is as follows.
In Sec.\,\ref{Lf} we describe how the interactions of the new particles, namely the scalar LQs and
the light RH neutrinos, with SM fermions can affect the \,$s\to d\slashed E$\, operators.
In Sec.\,\ref{amps} we derive the amplitudes for the hyperon and kaon decay modes of interest
and formulate their rates.
In Sec.\,\ref{constr} we deal with the pertinent limitations on the Yukawa couplings of
the LQs and present our numerical results.
To illustrate the potential impact of the scalar LQs on these hyperon and kaon processes, we
discuss two distinct cases.
In the first one, only SM quarks and leptons take part in the Yukawa interactions of the LQs.
In the second case, we let the RH neutrinos couple directly to the LQs as well as the quarks.
In each of these instances, it is necessary to have at least two different scalar LQs in order
to respect the existing \,$K\to\pi\nu\bar\nu$\, data and raise the hyperon rates significantly
at the same
time.\footnote{Invoking two scalar LQs to suppress certain processes and enhance others has
previously been employed to address the $B$ anomalies, such as
in \cite{Crivellin:2017zlb,Buttazzo:2017ixm}.}
We show that these two scenarios can furnish very dissimilar sets of predictions
for the hyperon rates which may be experimentally tested.
Part of the reason for the dissimilarity is that due to gauge invariance the SM neutrinos belong
to weak doublets along with the charged leptons, which causes the first scenario to be subject
to more restraints than the second.
We will also comment briefly on how our findings may have to be altered in order to accommodate
the $B$ anomalies if they turn out to proceed from NP.
In Sec.\,\ref{concl} we draw our conclusions.

\section{Leptoquark interactions\label{Lf}}

Among LQs which can have renormalizable couplings to SM fermions without violating baryon- and
lepton-number conservations and SM gauge symmetries, there are three which have spin 0 and can
at tree level induce FCNC transitions with missing energy among down-type
quarks~\cite{Dorsner:2016wpm}.
Following the nomenclature of Ref.\,\cite{Dorsner:2016wpm}, we denote these scalar LQs, with their
assignments under the SM gauge groups SU(3$)_{\rm color}\times{\rm SU}(2)_L\times{\rm U}(1)_Y$,
by $S_1\, \big(\bar 3,1,1/3\big)$, $\tilde R_2\, (3,2,1/6)$, and $S_3\, \big(\bar 3,3,1/3\big)$.
From this point on, we concentrate on the contributions of the first two.
In terms of their components,
\begin{align}
S_1^{} & \,=\, S_1^{1/3} \,, &
\tilde R_2^{} & \,=\, \left(\begin{array}{c} \tilde R_2^{2/3} \medskip \\
\tilde R_2^{-1/3^{\vphantom{|}}} \end{array}\right) , ~~~ ~~~~
\end{align}
where the superscripts indicate their electric charges.

We incorporate into the theory three RH neutrinos, which are singlets under the SM gauge groups
and called $\texttt{\textsl N}_1$, $\texttt{\textsl N}_2$, and $\texttt{\textsl N}_3$.
We assume that they and the SM neutrinos are all of Dirac nature.
In addition, we suppose that $\texttt{\textsl N}_{1,2,3}$ have masses small enough to be
neglected in the hyperon and kaon processes of concern and are sufficiently long-lived that
they do not decay inside detectors.

We express the Lagrangian for the renormalizable interactions of $S_1$ and $\tilde R_2$
with SM fermions plus $\texttt{\textsl N}_{1,2,3}$ as
\begin{align} \label{Llq}
{\cal L}_{\textsc{lq}}^{} & \,=\, \Big( \texttt Y_{1,jy\,}^{\textsc{ll}}
\overline{q_j^{\rm c}}\, i\tau_2^{} l_y^{} + \tilde{\textsc y}_{1,jy\,}^{\textsc{rr}}
\overline{d_j^{\rm c}}\, \texttt{\textsl N}_y^{} \Big) S_1^{}
+ \texttt Y_{2,jy\,}^{\textsc{rl}} \overline{d_j^{}}\tilde R_2^{\textsc t\,} i\tau_2^{} l_y^{}
+ \tilde{\textsc y}_{2,jy\,}^{\textsc{lr}}\overline{q_j^{}}\tilde R_2^{}\texttt{\textsl N}_y^{}
\,+\, \rm H.c. \,,
\end{align}
where the $\texttt Y_{jy}$ and $\tilde{\textsc y}_{jy}$ are generally complex elements of
the LQ Yukawa matrices, summation over family indices \,$j,y=1,2,3$\, is implicit,
$q_j$ $(l_y)$ and $d_j$ symbolize a left-handed quark (lepton) doublet and right-handed
down-type quark singlet, respectively, and $\tau_2^{}$ stands for the second
Pauli matrix.\footnote{In Eq.\,(\ref{Llq}) we include only the minimal ingredients which
serve our purposes pertaining to the \,$s\to d\slashed E$\, reactions of interest.
In principle, one could have more Yukawa terms, such as
$\overline{u_j^{\rm c}}e_y^{}S_1^{}$, with $u_j$ ($e_y$) designating a right-handed
up-type quark (charged lepton) field, and/or introduce scalar couplings, such as
$H^\dagger S_1\tilde R_2$ involving the SM Higgs doublet $H$ leading to mixing between
the LQ components of the same electric charge~\cite{Hirsch:1996qy,Dorsner:2017wwn}.
Such extensions would be accompanied by new free parameters and contribute to processes beyond
the scope of this paper.
Therefore, we refrain from considering LQ interactions not contained in Eq.\,(\ref{Llq}).}
To obtain decay amplitudes from ${\cal L}_{\textsc{lq}}$, we need to write it in terms of
mass eigenstates.
For the processes of interest, it is convenient to do so in the mass basis of the down-type
fermions, in which case
\begin{align} \label{ql}
q_j^{} & \,=\, \left(\begin{array}{c} {\cal V}_{kj}^{*~} \texttt{\textsl U}_{kL}^{}
\medskip \\ \texttt{\textsl D}_{jL_{\vphantom{|}}}^{} \end{array}\right) , &
l_j^{} & \,=\, \left(\begin{array}{c} {\cal U}_{jk\,}^{} \nu_k^{} \medskip \\
\ell_{jL}^{} \end{array}\right) , &
d_j^{} & \,=\, \texttt{\textsl D}_{jR}^{} \,, ~~
\end{align}
where \,$k=1,2,3$\, is implicitly summed over, \,${\cal V}\equiv{\cal V}_{\textsc{ckm}}$
$({\cal U}\equiv{\cal U}_{\textsc{pmns}})$ is the Cabibbo-Kobayashi-Maskawa quark
(Pontecorvo-Maki-Nakagawa-Sakata neutrino) mixing matrix,
\,$f_{L,R}^{}=\frac{1}{2}(1\mp\gamma_5^{})f$, and \,$\texttt{\textsl U}_{1,2,3}$ $(=u,c,t)$,\,
$\texttt{\textsl D}_{1,2,3}$ $(=d,s,b)$,\, $\nu_{1,2,3}^{}$,\, and
\,$\ell_{1,2,3}$ $(=e,\mu,\tau)$\, represent mass eigenstates.
Since the left-handed neutrinos' masses are tiny, we can work with
\,$\nu_{\ell_j}={\cal U}_{jk}^{}\nu_k^{}$\, instead of $\nu_j^{}$.
Thus, Eq.\,(\ref{Llq}) becomes
\begin{align} \label{Llqs}
{\cal L}_{\textsc{lq}}^{} & \,=\,
\Big\{ \texttt Y_{1,jy}^{\textsc{ll}} \Big[ {\cal V}_{kj}^*
\overline{(\texttt{\textsl U}_{kL})\raisebox{1pt}{$^{\rm c}$}} \ell_{yL}^{}
- \overline{(\texttt{\textsl D}_{jL})\raisebox{1pt}{$^{\rm c}$}} \nu_{\ell_y}^{} \Big]
+ \tilde{\textsc y}_{1,jy\,}^{\textsc{rr}}
\overline{(\texttt{\textsl D}_{jR})\raisebox{1pt}{$^{\rm c}$}}\,
\texttt{\textsl N}_y^{} \Big\} S_1^{}
\nonumber \\ & ~~~~ +\,
\texttt Y_{2,jy}^{\textsc{rl}}\, \overline{\texttt{\textsl D}_{jR}^{}} \Big( \ell_{yL}^{}
\tilde R_2^{2/3} - \nu_{\ell_y}^{} \tilde R_2^{-1/3} \Big)
+ \tilde{\textsc y}_{2,jy}^{\textsc{lr}} \Big(
{\cal V}_{kj\,}^{} \overline{\texttt{\textsl U}_{kL}^{}} \tilde R_2^{2/3}
+ \overline{\texttt{\textsl D}_{jL}^{}} \tilde R_2^{-1/3} \Big) \texttt{\textsl N}_y^{}
\nonumber \\ & ~~~~ +\, \rm H.c.
\end{align}

These couplings allow $S_1$ and $\tilde R_2$ to mediate
$ds_{\;}\!\texttt{\textsl f}\texttt{\textsl f}'$ interactions at tree level,
the $\texttt{\textsl f}\texttt{\textsl f}'$ pair being unobserved,
with \,$\texttt{\textsl f}=\nu$ or $\texttt{\textsl N}$.\,
Assuming that both of the LQs are heavy, we find that ${\cal L}_{\textsc{lq}}$ gives rise to
the low-energy effective Lagrangian
\begin{align} \label{Ldsff'}
-{\cal L}_{ds\texttt{\textsl f}\texttt{\textsl f}'}^{} \,=\,
\overline{d}\gamma^\eta s\; \overline{\texttt{\textsl f}} \gamma_\eta^{} \big(
\texttt C_{\texttt{\textsl f}\texttt{\textsl f}'}^{\texttt V}
+ \gamma_5^{}\texttt C_{\texttt{\textsl f}\texttt{\textsl f}'}^{\texttt A} \big) \texttt{\textsl f}'
+ \overline{d} \gamma^\eta\gamma_5^{} s\; \overline{\texttt{\textsl f}} \gamma_\eta^{} \big(
\tilde{\textsf c}_{\texttt{\textsl f}\texttt{\textsl f}'}^{\textsc v}
+ \gamma_5^{} \tilde{\textsf c}{}_{\texttt{\textsl f}\texttt{\textsl f}'}^{\textsc a} \big)
\texttt{\textsl f}'  \,+\, {\rm H.c.} \,, &
\end{align}
the pertinent coefficients being given by
\begin{subequations} \label{Cff'}
\begin{align} \label{Cff'a}
\texttt C_{\nu\nu'}^{\texttt V} & =\, -\texttt C_{\nu\nu'}^{\texttt A} =\,
\frac{-\texttt Y_{1,1x}^{\textsc{ll}*}  \texttt Y_{1,2y}^{\textsc{ll}}}{8 m_{S_1}^2}
+ \frac{\texttt Y_{2,2x}^{\textsc{rl}*} \texttt Y_{2,1y}^{\textsc{rl}}}{8 m_{\tilde R_2}^2} \,, &
\tilde{\textsf c}_{\nu\nu'}^{\textsc v} & =\, -\tilde{\textsf c}_{\nu\nu'}^{\textsc a} =\,
  \frac{\texttt Y_{1,1x}^{\textsc{ll}*} \texttt Y_{1,2y}^{\textsc{ll}}}{8 m_{S_1}^2}
+ \frac{\texttt Y_{2,2x}^{\textsc{rl}*} \texttt Y_{2,1y}^{\textsc{rl}}}{8 m_{\tilde R_2}^2} \,, ~~~
\end{align}
\vspace{-3ex}
\begin{align} \label{Cff'b}
\texttt C_{\texttt{\textsl N}\texttt{\textsl N}'}^{\texttt V} & =\,
\texttt C_{\texttt{\textsl N}\texttt{\textsl N}'}^{\texttt A} =\,
\frac{-\tilde{\textsc y}_{1,1x}^{\textsc{rr}*}
\tilde{\textsc y}_{1,2y}^{\textsc{rr}}}{8 m_{S_1}^2}
+ \frac{\tilde{\textsc y}_{2,2x}^{\textsc{lr}*}
\tilde{\textsc y}_{2,1y}^{\textsc{lr}}}{8 m_{\tilde R_2}^2} \,, & ~
\tilde{\textsf c}_{\texttt{\textsl N}\texttt{\textsl N}'}^{\textsc v} & =\,
\tilde{\textsf c}_{\texttt{\textsl N}\texttt{\textsl N}'}^{\textsc a} =\,
\frac{-\tilde{\textsc y}_{1,1x}^{\textsc{rr}*}
\tilde{\textsc y}_{1,2y}^{\textsc{rr}}}{8 m_{S_1}^2}
- \frac{\tilde{\textsc y}_{2,2x}^{\textsc{lr}*}
\tilde{\textsc y}_{2,1y}^{\textsc{lr}}}{8 m_{\tilde R_2}^2} \,, ~
\end{align}
\end{subequations}
where the indices $x$ and $y$ label the lepton or $\texttt{\textsl N}$ species.

In the remainder of this article, we explore how these LQ-induced operators may affect
especially the aforementioned hyperon decays with missing energy as well as
\,$K\to\pi\pi'\slashed E$.\,
However, we will also touch on how our treatment might need to be adjusted in order
to accommodate the $B$-physics anomalies should future data establish them to be caused by NP.

\section{Amplitudes and rates\label{amps}}

In the hyperon sector ${\cal L}_{ds\texttt{\textsl f}\texttt{\textsl f}'}$ in Eq.\,(\ref{Ldsff'})
brings about the decay modes
\,$\mathfrak B\to\mathfrak B'\texttt{\textsl f}\bar{\texttt{\textsl f}}{}'$\,
involving the pairs of baryons
\,$\mathfrak{BB}'=\Lambda n,\Sigma^+p,\Xi^0\Lambda,\Xi^0\Sigma^0,\Xi^-\Sigma^-$,\, all having
spin 1/2, and \,$\Omega^-\to\Xi^-\texttt{\textsl f}\bar{\texttt{\textsl f}}{}'$,\,
where $\Omega^-$ has spin 3/2.
To deal with the amplitudes for these decays, we need the baryonic matrix elements of the quark
bilinears in Eq.\,(\ref{Ldsff'}) which can be estimated with the aid of chiral perturbation
theory at leading order \cite{He:2005we,Tandean:2019tkm}.
For \,$\mathfrak B\to\mathfrak B'\texttt{\textsl f}\bar{\texttt{\textsl f}}{}'$\,
the results are \cite{Tandean:2019tkm}
\begin{align} \label{<B'B>}
\big\langle\mathfrak B'\big|\overline{d}\gamma^\eta s\big|\mathfrak B\big\rangle & =\,
{\cal V}_{\mathfrak B'\mathfrak B}^{}\,\bar u_{\mathfrak B'}^{}\gamma^\eta u_{\mathfrak B}^{} \,,
\nonumber \\
\big\langle{\mathfrak B'}\big|\overline{d}\gamma^\eta\gamma_5^{}s\big|{\mathfrak B}\big\rangle
& =\, {\cal A}_{\mathfrak B'\mathfrak B}^{}\, \bar u_{\mathfrak B'}^{} \bigg( \gamma^\eta
- \frac{m_{\mathfrak B'}^{}+m_{\mathfrak B}^{}}{m_K^2-\texttt Q^2}\, \texttt Q^\eta{} \bigg)
\gamma_5^{} u_{\mathfrak B}^{} \,, ~~
\end{align}
where
\,${\cal V}_{\mathfrak B'\mathfrak B}=-3/\sqrt{6},-1,3/\sqrt{6},-1/\sqrt2,1$\, and
\,${\cal A}_{\mathfrak B'\mathfrak B}=-(D+3F)/\sqrt6,D-F,(3F-D)/\sqrt6$, $-(D+F)/\sqrt2,D+F$\,
for \,${\mathfrak B}'{\mathfrak B}=n\Lambda,p\Sigma^+,\Lambda\Xi^0,\Sigma^0\Xi^0,\Sigma^-\Xi^-$,\,
respectively, $\bar u_{\mathfrak B'}$ and $u_{\mathfrak B}$ represent the Dirac spinors of
the baryons, and \,$\texttt Q=p_{\mathfrak B}^{}-p_{\mathfrak B'}^{}$,\, with $p_{\mathfrak B}^{}$
and $p_{\mathfrak B'}^{}$ being their momenta.
For \,$\Omega^-\to\Xi^-\texttt{\textsl f}\bar{\texttt{\textsl f}}{}'$\, we
have~\cite{Tandean:2019tkm}
\begin{align} \label{<XO>}
\langle\Xi^-\big|\overline{d}\gamma^\eta\gamma_5^{}s|\Omega^-\rangle & \,=\, {\cal C}\,
\bar u_{\Xi}^{} \Bigg( u_{\Omega}^\eta + \frac{\tilde{\textsc q}{}^\eta\,
\tilde{\textsc q}{}_\kappa^{}}{m_K^2-\tilde{\textsc q}{}^2}\, u_\Omega^\kappa \Bigg) ,
\end{align}
where \,$\tilde{\textsc q}=p_{\Omega^-}^{}-p_{\Xi^-}^{}$\, and
$u_\Omega^\eta$ is a Rarita-Schwinger spinor.
The constants $D$, $F$, and $\cal C$ above occur in the lowest-order
chiral Lagrangian and can be fixed from baryon data.
In numerical work, we will adopt the same values of these and other input parameters
as those given in Ref.\,\cite{Tandean:2019tkm}.

With Eqs.\,\,(\ref{<B'B>}) and (\ref{<XO>}), we derive the amplitudes for
\,$\mathfrak B\to\mathfrak B'\texttt{\textsl f}\bar{\texttt{\textsl f}}{}'$\, and
\,$\Omega^-\to\Xi^-\texttt{\textsl f} \bar{\texttt{\textsl f}}{}'$.\,
Since in the scenarios under consideration the $\texttt{\textsl f}$ and $\texttt{\textsl f}'$
masses are assumed to be small compared to the pion mass, we then arrive at
the differential rates \cite{Tandean:2019tkm}
\begin{align} \label{G'B2B'ff'}
\frac{d\Gamma_{\mathfrak B\to\mathfrak B'\texttt{\textsl f}\bar{\texttt{\textsl f}}{}'}}{d\hat s}
\,=\, \frac{\lambda_{\mathfrak{BB}'}^{1/2}}{64 \pi^3 m_{\mathfrak B}^3} &
\bigg\{ \bigg[ \frac{\lambda_{\mathfrak{BB}'}}{3} + (m_{\mathfrak B}-m_{\mathfrak B'}^{})^2 \hat s
- \hat s^2 \bigg] \Big( \big|{\texttt C}_{\texttt{\textsl f}\texttt{\textsl f}'}^{\texttt V}\big|^2
+ \big|{\texttt C}_{\texttt{\textsl f}\texttt{\textsl f}'}^{\texttt A}\big|^2 \Big)
{\cal V}_{\mathfrak B'\mathfrak B}^2
\nonumber \\ & +
\bigg[ \frac{\lambda_{\mathfrak{BB}'}}{3} + (m_{\mathfrak B}+m_{\mathfrak B'}^{})^2 \hat s
- \hat s^2 \bigg] \Big(
\big|\tilde{\textsf c}_{\texttt{\textsl f}\texttt{\textsl f}'}^{\textsc v}\big|^2
+ \big|\tilde{\textsf c}_{\texttt{\textsl f}\texttt{\textsl f}'}^{\textsc a}\big|^2
\Big) {\cal A}_{\mathfrak B'\mathfrak B}^2 \bigg\} \,, ~~
\end{align}
\begin{align} \label{G'O2Xff'}
\frac{d\Gamma_{\Omega^-\to\Xi^-\texttt{\textsl f}\bar{\texttt{\textsl f}}{}'}}{d\hat s} \,=\,
\frac{\lambda_{\Omega^-\Xi^-\,}^{1/2} {\cal C}^2}{768 \pi^{3\,} m_{\Omega^-}^3} &
\Bigg( \frac{\lambda_{\Omega^-\Xi^-}}{3m_{\Omega^-}^2} + 4\hat s \Bigg)^{\vphantom{\int_\int^\int}}
\big[ (m_{\Omega^-}+m_{\Xi^-})^2-\hat s \big] \Big(
\big|\tilde{\textsf c}_{\texttt{\textsl f}\texttt{\textsl f}'}^{\textsc v}\big|^2
+ \big|\tilde{\textsf c}_{\texttt{\textsl f}\texttt{\textsl f}'}^{\textsc a}\big|^2 \Big) \,,
\end{align}
where \,$\hat s=(p_{\texttt{\textsl f}}^{}+p_{\texttt{\textsl f}'}^{})^2$\, and
\,$\lambda_{XY}^{}=m_X^4-2\big(m_Y^2+\hat s\big)m_X^2+\big(m_Y^2-\hat s\big)\raisebox{1pt}{$^2$}$.\,
As in Ref.\,\cite{Tandean:2019tkm}, in our numerical evaluation of the hyperon rates we include
form-factor effects not yet taken into account in Eqs.\,\,(\ref{<B'B>}) and (\ref{<XO>}).
Particularly, in Eq.\,(\ref{G'B2B'ff'}) we apply the changes
\,${\cal V}_{\mathfrak B'\mathfrak B} \to \big(1+2\hat s/M_V^2\big)
{\cal V}_{\mathfrak B'\mathfrak B}$\,
and
\,${\cal A}_{\mathfrak B'\mathfrak B} \to \big(1+2\hat s/M_A^2\big)
{\cal A}_{\mathfrak B'\mathfrak B}$\,
with \,$M_V=0.97(4)$\,GeV and \,$M_A=1.25(15)$\,GeV,\, in line with the parametrization
commonly employed in experimental analyses of the charged-current semileptonic decays of
hyperons \cite{Bourquin:1981ba,Gaillard:1984ny,Hsueh:1988ar,Dworkin:1990dd,Batley:2006fc}.
Moreover, in Eq.\,(\ref{G'O2Xff'}) we modify $\cal C$ to
\,${\cal C}/\big(1-\hat s/M_A^2\big)\raisebox{1pt}{$^2$}$.\,

In the kaon sector, ${\cal L}_{ds\texttt{\textsl f}\texttt{\textsl f}'}$ in Eq.\,(\ref{Ldsff'})
is consequential for \,$K\to\pi\texttt{\textsl f}\bar{\texttt{\textsl f}}{}'$\, and
\,$K\to\pi\pi'\texttt{\textsl f}\bar{\texttt{\textsl f}}{}'$\, but not for
\,$K\to\texttt{\textsl f}\bar{\texttt{\textsl f}}{}'$\, which is helicity suppressed because in
this study we focus on the
\,$m_{\texttt{\textsl f},\texttt{\textsl f}'}\simeq0$\, case.
It follows that the relevant mesonic matrix elements are
\begin{align}
\langle\pi^-(p_\pi^{})|\bar d\gamma^\eta s|K^-(p_K^{})\rangle & \,=\, p_K^\eta+p_\pi^\eta \,,
\nonumber \\
\langle\pi^0(p_0^{})\,\pi^-(p_-^{})|\bar d \gamma^\eta\gamma_5^{}s|K^-\rangle
& \,=\, \frac{i\sqrt2}{f_K^{}} \Bigg[ \big( p_0^\eta - p_-^\eta \big)
+ \frac{\big(p_0^\mu-p_-^\mu\big)\tilde{\texttt{\textsl q}}{}_\mu^{}\,
\tilde{\texttt{\textsl q}}{}^\eta}{m_K^2-\tilde{\texttt{\textsl q}}{}^2} \Bigg] \,, &
\nonumber \\
\langle\pi^0(p_1^{})\,\pi^0(p_2^{})|\bar d \gamma^\eta\gamma_5^{}s|\bar K^0\rangle
& \,=\, \frac{i}{f_K^{}} \Bigg[ \big( p_1^\eta + p_2^\eta \big)
+ \frac{\big(p_1^\mu+p_2^\mu\big)\tilde{\texttt{\textsl q}}{}_\mu^{}\,
\tilde{\texttt{\textsl q}}{}^\eta}{m_K^2-\tilde{\texttt{\textsl q}}{}^2} \Bigg] \,, &
\end{align}
where \,$f_K^{}=155.6(4)$\,MeV \cite{Tanabashi:2018oca} is the kaon decay constant,
\,$\tilde{\texttt{\textsl q}}=p_{K^-}-p_0^{}-p_-^{}=p_{\bar K^0}-p_1^{}-p_2^{}$,\, and
we have ignored form-factor effects.
Assuming isospin symmetry and making use of charge conjugation, we also have the relations
\,$\big\langle\pi^0\big|\bar d\gamma^\eta s\big|\bar K^0\big\rangle =
-\big\langle\pi^-\big|\bar d\gamma^\eta s\big|K^-\big\rangle/\sqrt2 =
-\big\langle\pi^0\big|\bar s\gamma^\eta d\big|K^0\big\rangle$\,
and
\,$\langle\pi^0\pi^0|\bar s \gamma^\eta\gamma_5^{}d|K^0\rangle =
\langle\pi^0\pi^0|\bar d \gamma^\eta\gamma_5^{}s|\bar K^0\rangle$.\,
Hence the amplitudes for \,$K^-\to\pi^-(\pi^0)\texttt{\textsl f}\bar{\texttt{\textsl f}}{}'$\,
and \,$K_L\to\pi^0(\pi^0)\texttt{\textsl f}\bar{\texttt{\textsl f}}{}'$\, with the approximation 
\,$\sqrt2 K_L=K^0+\bar K^0$\, are  \cite{Tandean:2019tkm,He:2019xxp}
\begin{align} \label{MK2pff}
{\cal M}_{K^-\to\pi^-\texttt{\textsl f}\bar{\texttt{\textsl f}}{}'} & \,=\, 2\,
\bar u_{\texttt{\textsl f}}^{}\, \slashed p_{\!K} \big(
\texttt C_{\texttt{\textsl f}\texttt{\textsl f}'}^{\texttt V}
+ \gamma_5^{} \texttt C_{\texttt{\textsl f}\texttt{\textsl f}'}^{\texttt A} \big)
v_{\bar{\texttt{\textsl f}}{}'}^{} \,,
\nonumber \\
{\cal M}_{K_L\to\pi^0\texttt{\textsl f}\bar{\texttt{\textsl f}}{}'} & \,=\,
\bar u_{\texttt{\textsl f}}^{}\, \slashed p_{\!K} \big[
\texttt C_{\texttt{\textsl f}'\texttt{\textsl f}}^{\texttt V*}
- \texttt C_{\texttt{\textsl f}\texttt{\textsl f}'}^{\texttt V}
+ \gamma_5^{} \big( \texttt C_{\texttt{\textsl f}'\texttt{\textsl f}}^{\texttt A*}
- \texttt C_{\texttt{\textsl f}\texttt{\textsl f}'}^{\texttt A} \big) 
\big]_{\vphantom{|_|^|}}^{\vphantom{|_|^|}} v_{\bar{\texttt{\textsl f}}{}'}^{} \,,  
\\ \label{MK2ppff}
{\cal M}_{K^-\to\pi^0\pi^-\texttt{\textsl f}\bar{\texttt{\textsl f}}{}'}^{} & \,=\,
\frac{i\sqrt2^{\vphantom{|_|^|}}}{f_K^{}}\, \bar u_{\texttt{\textsl f}}^{} 
\big( \slashed p{}_0^{} - \slashed p{}_-^{} \big) \big(
\tilde{\textsf c}_{\texttt{\textsl f}\texttt{\textsl f}'}^{\textsc v}
+ \gamma_5^{} \tilde{\textsf c}_{\texttt{\textsl f}\texttt{\textsl f}'}^{\textsc a} \big)
v_{\bar{\texttt{\textsl f}}{}'}^{} \,,
\nonumber \\
{\cal M}_{K_L\to\pi^0\pi^0\texttt{\textsl f}\bar{\texttt{\textsl f}}{}'}^{} & \,=\,
\frac{i}{\sqrt2\,f_K^{}}\, \bar u_{\texttt{\textsl f}}^{} \big(\slashed p{}_1^{} + \slashed p{}_2^{}
\big) \big[ \tilde{\textsf c}_{\texttt{\textsl f}'\texttt{\textsl f}}^{\textsc v*}
+ \tilde{\textsf c}_{\texttt{\textsl f}\texttt{\textsl f}'}^{\textsc v}
+ \gamma_5^{} \big( \tilde{\textsf c}_{\texttt{\textsl f}'\texttt{\textsl f}}^{\textsc a*}
+ \tilde{\textsf c}_{\texttt{\textsl f}\texttt{\textsl f}'}^{\textsc a} \big)
\big] v_{\bar{\texttt{\textsl f}}}^{} ~~
\end{align}
in the \,$m_{\texttt{\textsl f},\texttt{\textsl f}'}=0$\, limit.
The expressions in Eq.\,(\ref{MK2pff}) lead to the differential decay rates
\begin{align} \label{GK2pff}
\frac{d\Gamma_{K^-\to\pi^-\texttt{\textsl f}\bar{\texttt{\textsl f}}{}'}^{}}{d\hat s} & \,=\,
\frac{\lambda_{K^+\pi^+\,}^{3/2}}{192\pi^3 m_{K^+}^3}
\Big( \big|\texttt C_{\texttt{\textsl f}\texttt{\textsl f}'}^{\texttt V}\big|^2
+ \big|\texttt C_{\texttt{\textsl f}\texttt{\textsl f}'}^{\texttt A}\big|^2 \Big) \,,
\nonumber \\
\frac{d\Gamma_{K_L^{}\to\pi^0\texttt{\textsl f}\bar{\texttt{\textsl f}}{}'}^{}}{d\hat s} & \,=\,
\frac{\lambda_{K^0\pi^0\,}^{3/2}}{768\pi^3 m_{K^0}^3} \Big(
\big| \texttt C_{\texttt{\textsl f}'\texttt{\textsl f}}^{\texttt V*}
- \texttt C_{\texttt{\textsl f}\texttt{\textsl f}'}^{\texttt V} \big|^2
+ \big| \texttt C_{\texttt{\textsl f}'\texttt{\textsl f}}^{\texttt A*}
- \texttt C_{\texttt{\textsl f}\texttt{\textsl f}'}^{\texttt A} \big|^2 \Big) \,.
\end{align}
From Eq.\,(\ref{MK2ppff}), we arrive at the double-differential decay rates
\begin{align} \label{G''K2ppff'}
\frac{d^2\Gamma_{K^-\to\pi^0\pi^-\texttt{\textsl f}\bar{\texttt{\textsl f}}{}'}}
{d\hat s\,d\hat\varsigma} \,=~ & \frac{\beta_{\hat\varsigma}^3\, \tilde\lambda{}_{K^-}^{1/2}}
{18(4\pi)^5 f_K^2 m_{K^-}^3} \raisebox{1.7pt}{$\Big($} \tilde\lambda_{K^-}^{}
+ 12 \hat s\hat\varsigma    \raisebox{1.7pt}{$\Big)$} \Big(
\big|\tilde{\textsf c}_{\texttt{\textsl f}\texttt{\textsl f}'}^{\textsc v}\big|^2
+ \big|\tilde{\textsf c}_{\texttt{\textsl f}\texttt{\textsl f}'}^{\textsc a}\big|^2 \Big) \,,
\nonumber \\
\frac{d^2\Gamma_{K_L\to\pi^0\pi^0\texttt{\textsl f}\bar{\texttt{\textsl f}}{}'}}
{d\hat s\,d\hat\varsigma} \,=~ &
\frac{\beta_{\hat\varsigma}^{}\, \tilde\lambda{}_{K^0}^{3/2}}{48 (4\pi)^5 f_K^2 m_{K^0}^3}
\Big( \big| \tilde{\textsf c}_{\texttt{\textsl f}'\texttt{\textsl f}}^{\textsc v*}
+ \tilde{\textsf c}_{\texttt{\textsl f}\texttt{\textsl f}'}^{\textsc v} \big|^2
+ \big| \tilde{\textsf c}_{\texttt{\textsl f}'\texttt{\textsl f}}^{\textsc a*}
+ \tilde{\textsf c}_{\texttt{\textsl f}\texttt{\textsl f}'}^{\textsc a} \big|^2 \Big) \,,
\end{align}
where
\begin{align}
\beta_{\hat\varsigma}^{} \,=\, \sqrt{1-\frac{4m_\pi^2}{\hat\varsigma}} \,, ~~~~~
\hat\varsigma \,=\, (p_0^{}+p_-^{})^2 & \,=\, (p_1^{}+p_2^{})^2 \,, ~~~~~
\tilde\lambda_K^{} \,=\, \big(m_K^2-\hat s-\hat\varsigma\big)^2
-4 \hat s \hat\varsigma \,.
\end{align}
The $\hat s$ and $\hat\varsigma$ integration ranges are \,$0\le\hat s\le(m_{K^-,K^0}-2m_\pi)^2$\,
and \,$4m_\pi^2\le\hat\varsigma\le\big(m_{K^-,K^0}^{}-\hat s{}^{1/2}\big)\raisebox{1pt}{$^2$}$\,
for the $K^-$ and $K_L$ channels, respectively.

If \,$\texttt{\textsl f}'\neq\texttt{\textsl f}$,\, the extra channel
$K_L\to\pi^0(\pi^0)\texttt{\textsl f}'\bar{\texttt{\textsl f}}$ also occurs, whose rate formula is
the same as $\Gamma_{K_L\to\pi^0(\pi^0)\texttt{\textsl f}\bar{\texttt{\textsl f}}{}'}$
but with $\texttt{\textsl f}$ and $\texttt{\textsl f}'$ interchanged.
For \,$\texttt{\textsl f}'\neq\texttt{\textsl f}$,\, if
$\texttt C_{\texttt{\textsl f}'\texttt{\textsl f}}^{\texttt V,\texttt A}$ and
$\tilde{\textsf c}_{\texttt{\textsl f}'\texttt{\textsl f}}^{\textsc v,\textsc a}$ are not zero,
they are generally independent from
$\texttt C_{\texttt{\textsl f}\texttt{\textsl f}'}^{\texttt V,\texttt A}$ and
$\tilde{\textsf c}_{\texttt{\textsl f}\texttt{\textsl f}'}^{\textsc v,\textsc a}$
and also induce \,$K^-\to\pi^-(\pi^0)\texttt{\textsl f}'\bar{\texttt{\textsl f}}$\, as well as
\,$\mathfrak B\to\mathfrak B'\texttt{\textsl f}'\bar{\texttt{\textsl f}}$\, and
\,$\Omega^-\to\Xi^-\texttt{\textsl f}'\bar{\texttt{\textsl f}}$.\,

In the instances we discuss in the next section, the emitted fermions are either SM neutrinos
with different flavors or the RH neutrinos, implying that the SM and NP amplitudes do not
interfere in the decay rates.
Furthermore, the SM contributions to the hyperon modes and \,$K\to\pi\pi'\slashed E$\, are
significantly smaller than the currently allowed room for potential NP in these
processes~\cite{Tandean:2019tkm} and hence can be neglected.

\section{Constraints and numerical results\label{constr}}

For definiteness and simplicity, we look at a couple of distinct situations as examples.
In the first one (A) the RH neutrinos, $\texttt{\textsl N}_{1,2,3}$, are absent.
In the second example (B) they are present and couple to the LQs and quarks.
Numerically, we will treat the nonzero Yukawa couplings phenomenologically, letting one or
more of them vanish or differ substantially from the others, in order to comply with various
constraints.

Before proceeding, we integrate the differential rates of the baryon and kaon decays
from the previous section to obtain the branching fractions
\begin{align} \label{BB2B'ff'}
{\cal B}\big(\Lambda\to n\texttt{\textsl f}\bar{\texttt{\textsl f}}{}'\big) & \,=\,
\Big( 8.89\, \big|\texttt C_{\texttt{\textsl f}\texttt{\textsl f}'}^{\texttt V}\big|^2
+ 14.1\, \big|\tilde{\textsf c}_{\texttt{\textsl f}\texttt{\textsl f}'}^{\textsc v}\big|^2
\Big) 10^7 \rm\;GeV^4 \,, &
\nonumber \\
{\cal B}\big(\Sigma^+\to p\texttt{\textsl f}\bar{\texttt{\textsl f}}{}'\big) & \,=\,
\Big( 10.2\, \big|\texttt C_{\texttt{\textsl f}\texttt{\textsl f}'}^{\texttt V}\big|^2
+ 3.67\, \big|\tilde{\textsf c}_{\texttt{\textsl f}\texttt{\textsl f}'}^{\textsc v}\big|^2
\Big) 10^7 \rm\;GeV^4 \,,
\nonumber \\
{\cal B}\big(\Xi^0\to\Lambda\texttt{\textsl f}\bar{\texttt{\textsl f}}{}'\big) & \,=\,
\Big( 18.5\, \big|\texttt C_{\texttt{\textsl f}\texttt{\textsl f}'}^{\texttt V}\big|^2
+ 1.99\, \big|\tilde{\textsf c}_{\texttt{\textsl f}\texttt{\textsl f}'}^{\textsc v}|^2
\Big) 10^7 \rm\;GeV^4 \,,
\nonumber \\
{\cal B}\big(\Xi^0\to\Sigma^0\texttt{\textsl f}\bar{\texttt{\textsl f}}{}'\big) & \,=\,
\Big( 0.57\, \big|\texttt C_{\texttt{\textsl f}\texttt{\textsl f}'}^{\texttt V}\big|^2
+ 2.75\, \big|\tilde{\textsf c}_{\texttt{\textsl f}\texttt{\textsl f}'}^{\textsc v}\big|^2
\Big) 10^7 \rm\;GeV^4 \,,
\nonumber \\
{\cal B}\big(\Xi^-\to\Sigma^-\texttt{\textsl f}\bar{\texttt{\textsl f}}{}'\big) & \,=\,
\Big( 0.70\, \big|\texttt C_{\texttt{\textsl f}\texttt{\textsl f}'}^{\texttt V}\big|^2
+ 3.38\, \big|\tilde{\textsf c}_{\texttt{\textsl f}\texttt{\textsl f}'}^{\textsc v}\big|^2
\Big) 10^7 \rm\;GeV^4 \,,
\nonumber \\
{\cal B}\big(\Omega^-\to\Xi^-\texttt{\textsl f}\bar{\texttt{\textsl f}}{}'\big) & \,=\, 1.60\,
\big|\tilde{\textsf c}_{\texttt{\textsl f}\texttt{\textsl f}'}^{\textsc v}\big|^2 \times
10^9 \rm\;GeV^4 \,,
\end{align}
\begin{align} \label{BK2piff'}
{\cal B}\big(K^-\to\pi^-\texttt{\textsl f}\bar{\texttt{\textsl f}}{}'\big) & \,=\, 2.59\,
\big|\texttt C_{\texttt{\textsl f}\texttt{\textsl f}'}^{\texttt V}\big|^2 \times
10^{10} \rm\;GeV^4 \,,
\nonumber \\
{\cal B}\big(K_L^{}\to\pi^0\texttt{\textsl f}\bar{\texttt{\textsl f}}{}'\big) & \,=\, 2.91\,
\big| \texttt C_{\texttt{\textsl f}'\texttt{\textsl f}}^{\texttt V*}
- \texttt C_{\texttt{\textsl f}\texttt{\textsl f}'}^{\texttt V} \big|^2 \times
10^{10} \rm\;GeV^4 \,, &
\end{align}
\begin{align} \label{BK2pipi'ff'}
{\cal B}\big(K^-\to\pi^-\pi^0\texttt{\textsl f}\bar{\texttt{\textsl f}}{}'\big) & \,=\, 1.26\,
\big|\tilde{\textsf c}_{\texttt{\textsl f}\texttt{\textsl f}'}^{\textsc v}\big|^2 \times
10^6 \rm\;GeV^4 \,,
\nonumber \\
{\cal B}\big(K_L^{}\to\pi^0\pi^0\texttt{\textsl f}\bar{\texttt{\textsl f}}{}'\big) & \,=\, 4.24\,
\big| \tilde{\textsf c}_{\texttt{\textsl f}'\texttt{\textsl f}}^{\textsc v*}
+ \tilde{\textsf c}_{\texttt{\textsl f}\texttt{\textsl f}'}^{\textsc v} \big|^2 \times
10^6 \rm\;GeV^4 \,. &
\end{align}
These results already incorporate the relations
\,$|\texttt C_{\texttt{\textsl f}\texttt{\textsl f}'}^{\texttt A}| =
|\texttt C_{\texttt{\textsl f}\texttt{\textsl f}'}^{\texttt V}|$\,
and
\,$|\tilde{\textsf c}_{\texttt{\textsl f}\texttt{\textsl f}'}^{\textsc a}| =
|\tilde{\textsf c}_{\texttt{\textsl f}\texttt{\textsl f}'}^{\textsc v}|$\,
from Eq.\,(\ref{Cff'}).

\subsection{SM fermions only\label{smf}}

Among FCNC kaon decays with missing energy, the most precise data available to date are
the branching fractions~\cite{Tanabashi:2018oca}
\,${\mathcal B}(K^+\to\pi^+\nu\nu)_{\rm exp}^{}=1.7(1.1)\times10^{-10}$ and
\,${\mathcal B}\big(K_L\to\pi^0\nu\bar\nu\big)_{\rm exp}^{} < 3.0\times10^{-9}$
at 90\% confidence level~(CL).
Evidently, they are not very far from what the SM predicts \cite{Bobeth:2017ecx}:
\,${\mathcal B}(K^+\to\pi^+\nu\bar\nu)_{\textsc{sm}}^{} =
\big(8.5_{-1.2}^{+1.0}\big)\times10^{-11}$\, and
\,${\mathcal B}(K_L\to\pi^0\nu\bar\nu)_{\textsc{sm}}^{} =
\big(3.2_{-0.7}^{+1.1}\big)\times10^{-11}$.\,
The extent to which these experimental numbers deviate from their SM counterparts reflects
how much possible NP is permitted to influence these modes.
We may then adopt the upper ends of the 90\%-CL ranges of the deviations,
\begin{align} \label{K2pinv}
\Delta{\mathcal B}(K^-\to\pi^-\slashed E) & \,<\, 2.7\times10^{-10} \,, &
\Delta{\mathcal B}(K_L\to\pi^0\slashed E) & \,<\, 3.0\times10^{-9}  \,, &
\end{align}
as caps on the NP contributions.
In view of Eq.\,(\ref{GK2pff}), this applies to $\texttt C_{\nu\nu'}^{\texttt V,\texttt A}$
in Eq.\,(\ref{Cff'a}).
In the absence of $\texttt{\textsl N}_{1,2,3}$, we set
\,$\texttt C_{\texttt{\textsl N}\texttt{\textsl N}'}^{\texttt V,\texttt A} =
\tilde{\textsf c}_{\texttt{\textsl N}\texttt{\textsl N}'}^{\textsc v,\textsc a} = 0$.\,

In general, the two parts of \,$\texttt C_{\nu\nu'}^{\texttt V}=-\texttt C_{\nu\nu'}^{\texttt A}$\,
in Eq.\,(\ref{Cff'a}) do not cancel each other.
Since they also enter $\tilde{\textsf c}_{\nu\nu'}^{\textsc v,\textsc a}$ but with the opposite
relative sign, and since all of these couplings participate in the hyperon rates in
Eqs.\,\,(\ref{G'B2B'ff'}) and (\ref{G'O2Xff'}), one way to amplify the hyperon rates and satisfy
Eq.\,\,(\ref{K2pinv}) simultaneously is if
\,$\texttt C_{\nu\nu'}^{\texttt V,\texttt A}=0$,\, which implies
\begin{align} \label{tcnn'}
\tilde{\textsf c}_{\nu\nu'}^{\textsc v} & \,=\, -\tilde{\textsf c}_{\nu\nu'}^{\textsc a} \,=\,
\frac{\texttt Y_{1,1x}^{\textsc{ll}*}\texttt Y_{1,2y}^{\textsc{ll}}}{4 m_{S_1}^2} \,=\,
\frac{\texttt Y_{2,2x}^{\textsc{rl}*}\texttt Y_{2,1y}^{\textsc{rl}}}{4 m_{\tilde R_2}^2} \,.
\end{align}
Consequently, as can be deduced from Eqs.\,\,(\ref{BB2B'ff'})-(\ref{BK2pipi'ff'}),
the \,$K\to\pi\slashed E$\, decays are now due to the SM alone, whereas \,$K\to\pi\pi'\slashed E$\,
and the hyperon modes can still undergo LQ-induced modifications.\footnote{Studies focusing on
the hyperon modes affected by NP manifesting the same chiral structure as that of the SM operator,
namely with
\,$\texttt C_{\nu\nu'}^{\texttt V}=-\texttt C_{\nu\nu'}^{\texttt A} =
-\tilde{\textsf c}_{\nu\nu'}^{\textsc v}=\tilde{\textsf c}_{\nu\nu'}^{\textsc a}$\,
in Eq.\,(\ref{Ldsff'}), can be found in \cite{Hu:2018luj,Mahmood:2019bpk}.}
In the rest of this subsection, we investigate the extent to which the hyperon rates may be
enhanced with the choice in Eq.\,(\ref{tcnn'}) subject to additional relevant requirements.
However, we shall also briefly look at the \,$\texttt C_{\nu\nu'}^{\texttt V,\texttt A}\neq0$\,
case.

The interactions in Eq.\,(\ref{Llqs}) can impact the mixing of neutral kaons $K^0$ and $\bar K^0$
via four-quark operators which arise from box diagrams, with the LQs and SM leptons running around
the loops, and are given in the effective
Hamiltonian~\cite{Dorsner:2016wpm,Bobeth:2017ecx,Mandal:2019gff}
\begin{align} \label{Hds2}
{\cal H}_{|\Delta S|=2}^{\textsc{lq}} & \,=\, \frac{\big( \raisebox{0pt}{\small$\sum_x$}
\texttt Y_{1,1x}^{\textsc{ll}} \texttt Y_{1,2x}^{\textsc{ll}*} \big)\raisebox{1pt}{$\!^2$}}
{128\pi^2\, m_{S_1}^2}\, \bar s_L^{}\gamma^\eta d_L^{}\, \bar s_L^{}\gamma_\eta^{}d_L^{}
+ \frac{\big( \raisebox{0pt}{\small$\sum_x$} \texttt Y_{2,2x}^{\textsc{rl}}
\texttt Y_{2,1x}^{\textsc{rl}*} \big)\raisebox{1pt}{$\!^2$}}{64\pi^2\, m_{\tilde R_2}^2}\,
\bar s_R^{}\gamma^\eta d_R^{}\, \bar s_R^{}\gamma_\eta^{}d_R^{}
\,+\, {\rm H.c.}
\end{align}
Its matrix element between the $K^0$ and $\bar K^0$ states contributes to the kaon-mixing
parameters \,$\Delta m_K^{} \propto {\rm Re} M_{\bar KK}$\, and
\,$\epsilon_K^{} \propto {\rm Im} M_{\bar KK}$,\, where
\,$M_{\bar KK}=\langle\bar K^0|{\cal H}_{|\Delta S|=2}|K^0\rangle/(2m_{K^0})$.\,
Given that $\Delta m_K^{}$ and $\epsilon_K^{}$ have been well measured, the constraints
from them on the LQ effects are stringent~\cite{Mandal:2019gff}.
Nevertheless, we see from Eq.\,(\ref{Hds2}) that these restrictions can be evaded by assigning
the contributing elements of the first and second rows of $\texttt Y_1^{\textsc{ll}}$ and
$\texttt Y_2^{\textsc{rl}}$ to separate columns.

The interactions in Eq.\,(\ref{Llqs}) also give rise to lepton-flavor-violating $ds\ell\ell'$
operators in
\begin{align} \label{Ldsll'}
-{\cal L}_{ds\ell\ell'}^{} \,=\, \overline d\gamma^\eta s~ \overline{\ell} \gamma_\eta^{}
\big( \texttt V_{\ell\ell'}^{} + \gamma_5^{} \texttt A_{\ell\ell'}^{}\big) \ell'
+ \overline d\gamma^\eta\gamma_5^{}s\, \overline{\ell} \gamma_\eta^{} \big(
\tilde{\textsc v}_{\ell\ell'}^{} + \gamma_5^{} \tilde{\textsc a}_{\ell\ell'}^{} \big) \ell'
\,+\, {\rm H.c.} \,, &
\end{align}
where from tree-level LQ-exchange diagrams \cite{Dorsner:2016wpm,He:2019xxp,Mandal:2019gff}
\begin{align} \label{VV}
\texttt V_{\ell\ell'}^{} \,=\, -\texttt A_{\ell\ell'}^{} \,=\, \tilde{\textsc v}_{\ell\ell'}^{}
\,=\, -\tilde{\textsc a}_{\ell\ell'}^{} & \,=\,
\frac{\texttt Y_{2,2x}^{\textsc{rl}*}\texttt Y_{2,1y}^{\textsc{rl}}}{8 m_{\tilde R_2}^2} \,. ~~~~~
\end{align}
Accordingly, if the pair of the lepton-family indices of
\,$\texttt Y_{2,2x}^{\textsc{rl}*}\texttt Y_{2,1y}^{\textsc{rl}}$\, has the values \,$xy=12,21$,\,
corresponding to \,$\ell\ell'=e\mu,\mu e$,\, the $\overline d\gamma^\eta s$ and
$\overline d\gamma^\eta\gamma_5^{}s$ terms in Eq.\,(\ref{Ldsll'}), plus their H.c., will bring
about \,$K\to\pi e^\pm\mu^\mp$\, and \,$K_L\to e^\pm\mu^\mp$,\, respectively, both of which
have severe experimental restraints.
They can be eluded by ensuring that $xy$ is neither 12 nor 21.
These choices moreover do not lead to the one-loop diagrams causing the \,$\mu\to e\gamma,3e$\,
decays and \,$\mu\to e$\, transitions in nuclei and hence can escape the limits from their quests.

The conditions described in the preceding two paragraphs dictate that some of the elements of
the Yukawa matrices $\texttt Y_1^{\textsc{ll}}$ and $\texttt Y_2^{\textsc{rl}}$ be substantially
smaller than the others.
To avoid further constraints from processes involving the $b$ quark, we assume that the third-row
elements of these matrices are sufficiently tiny.
For the sake of simplicity, we can then approximate the matrices by setting their relatively
suppressed elements to zero.\footnote{This kind of treatment of LQ Yukawa matrices has been applied
to other contexts, such as in \cite{Dorsner:2016wpm,Cai:2017wry,Fajfer:2018bfj}.}

Thus, we can write down a sample set of the desired matrices as
\begin{align} \label{Y1Y2} &&
\texttt Y_1^{\textsc{ll}} & \,= \left(\begin{array}{ccc} 0 & 0 & y_{1,13}^{}
\smallskip \\ 0 & y_{1,22}^{} & 0 \smallskip \\ 0 & 0 & 0 \end{array}\right) , &
\texttt Y_2^{\textsc{rl}} & \,= \left(\begin{array}{ccc} 0 & y_{2,12}^{} & 0
\smallskip \\ 0 & 0 & y_{2,23}^{} \smallskip \\ 0 & 0 & 0 \end{array}\right) . && ~~~
\end{align}
Incorporating these into Eqs.\,\,(\ref{tcnn'}) and (\ref{VV}), we obtain
\begin{align} \label{vll'}
\tilde{\textsf c}_{\nu_\tau\nu_\mu}^{\textsc v} \,=\,
-\tilde{\textsf c}_{\nu_\tau\nu_\mu}^{\textsc a} \,=\,
\frac{y_{1,13}^*\, y_{1,22}^{}}{4 m_{S_1}^2} & \,=\,
\frac{y_{2,23}^*\, y_{2,12}^{}}{4 m_{\tilde R_2}^2} \,,
\\ \label{Vll'}
\texttt V_{\tau\mu}^{} \,=\, -\texttt A_{\tau\mu}^{} \,=\,
\tilde{\textsc v}_{\tau\mu}^{} \,=\, -\tilde{\textsc a}_{\tau\mu}^{} & \,=\,
\frac{y_{2,23}^*\, y_{2,12}^{}}{8 m_{\tilde R_2}^2} \,, ~~ ~~~
\end{align}
with the other coefficients in ${\cal L}_{ds\texttt{\textsl f}\texttt{\textsl f}'}$
and ${\cal L}_{ds\ell\ell'}$ in Eqs.\,\,(\ref{Ldsff'}) and (\ref{Ldsll'}) being zero.
Alternatively, we could pick one of three other sets, from which the analogous products of
nonzero matrix elements are given by
$\big(y_{1,12}^*\, y_{1,23}^{},\, y_{2,22}^*\, y_{2,13}^{}\big)$,
$\big(y_{1,11}^*\, y_{1,23}^{},\, y_{2,21}^*\, y_{2,13}^{}\big)$, and
$\big(y_{1,13}^*\, y_{1,21}^{},\, y_{2,23}^*\, y_{2,11}^{}\big)$, respectively.

If \,$xy=13,31,23,32$\, in Eq.\,(\ref{VV}), the operators in Eq.\,(\ref{Ldsll'}) generate
the lepton-flavor-violating \,$|\Delta S|=1$\, tau decays
\,$\tau\to\ell\bar K^{(*)},\ell K^{(*)},\ell\pi^\pm K^\mp$\, with \,$\ell=e,\mu$,\, and so
their empirical bounds are unavoidable by the selections of the Yukawa couplings in the previous
paragraph but way milder than the aforesaid restrictions on $e$-$\mu$ violation.
Numerically, among the $\tau$ decay constraints, the one on \,$\tau^-\to\mu^-\bar K^{*0}$\, turns
out to be the least demanding on the coefficients, as can be inferred from Ref.\,\cite{He:2019iqf},
and bears on the combinations already displayed in Eq.\,(\ref{Vll'}).
This translates into
\begin{subequations} \label{yylimit}
\begin{align} \label{y2y2limit}
\frac{\big|y_{2,23\,}^* y_{2,12}^{}\big|}{m_{\tilde R{}_2}^2} & \,<\, \frac{0.036}{\rm TeV^2}
\end{align}
and in view of Eq.\,(\ref{vll'})
\begin{align} \label{y1y1limit}
\frac{\big|y_{1,13\,}^* y_{1,22}^{}\big|}{m_{S_1}^2} & \,<\, \frac{0.036}{\rm TeV^2} \,.
\end{align}
\end{subequations}

Before moving on to the resulting hyperon predictions, we mention here additional potentially
relevant observables on which empirical information is available, but which at the moment entail
much weaker restraints than Eq.\,(\ref{yylimit}).
The $S_1$ parameters also participate in the decay \,$\tau\to\mu\gamma$\, induced by one-loop
diagrams~\cite{Dorsner:2016wpm}, but the rate turns out to be far below its current measured
bound because of loop suppression and the fact that the LQ interactions with the charged
leptons are chiral~\cite{Mandal:2019gff}, exclusively left-handed as Eq.\,(\ref{Llqs}) specifies.
Although the couplings of the $Z$ boson to a pair of SM leptons receive one-loop modifications
from the LQ interactions~\cite{Buttazzo:2017ixm}, the restrictions inferred from the existing $Z$
data are not yet competitive to Eq.\,(\ref{yylimit}).
Direct LQ searches at colliders have turned up negative to date~\cite{Tanabashi:2018oca,
Aaboud:2019bye,Aad:2020iuy,Sirunyan:2018kzh}, ruling out the lighter LQs, and most
recently ATLAS~\cite{Aaboud:2019bye,Aad:2020iuy} (CMS~\cite{Sirunyan:2018kzh}) has excluded
scalar LQs decaying fully into a quark and charged lepton (neutrino) at 95\%\,\,CL for masses
up to 1.8 (1.1) TeV, but this can be easily satisfied
by the parameters already fulfilling Eq.\,(\ref{yylimit}).

We can now put Eqs.\,\,(\ref{vll'}) and (\ref{yylimit}) together with the branching fractions of
the kaon decays \,$K\to\pi\pi'\texttt{\textsl f}\bar{\texttt{\textsl f}}{}'$\, and the hyperon
ones, \,$\mathfrak B\to\mathfrak B'\texttt{\textsl f}\bar{\texttt{\textsl f}}{}'$\, and
\,$\Omega^-\to\Xi^-\texttt{\textsl f}\bar{\texttt{\textsl f}}{}'$.\,
Thus, employing the formulas in Eqs.\,\,(\ref{BB2B'ff'}) and\,\,(\ref{BK2pipi'ff'}),
we predict the upper limits
\begin{align} \label{4body}
{\mathcal B}\big(K^-\to\pi^0\pi^-\nu_\tau\bar\nu_\mu\big) & \,<\, 1.0\times10^{-10} \,, &
{\mathcal B}\big(K_L\to\pi^0\pi^0\nu_\tau\nu_\mu\big) & \,<\, 6.9\times10^{-10} \,, ~~
\end{align}
\vspace{-1cm}
\begin{align} \label{B->B'ff'}
{\cal B}\big(\Lambda\to n\nu_\tau\bar\nu_\mu\big) & \,<\, 1.1\times10^{-8} \,, &
{\cal B}\big(\Sigma^+\to p\nu_\tau\bar\nu_\mu\big) & \,<\, 3.0\times10^{-9} \,,
\nonumber \\
{\cal B}\big(\Xi^0\to\Lambda\nu_\tau\bar\nu_\mu\big) & \,<\, 1.6\times10^{-9} \,, &
{\cal B}\big(\Xi^0\to\Sigma^0\nu_\tau\bar\nu_\mu\big) & \,<\, 2.2\times10^{-9} \,,
\nonumber \\
{\cal B}\big(\Xi^0\to\Sigma^0\nu_\tau\bar\nu_\mu\big) & \,<\, 2.7\times10^{-9} \,, &
{\cal B}\big(\Omega^-\to\Xi^-\nu_\tau\bar\nu_\mu\big) & \,<\, 1.3\times10^{-7} \,, ~~
\end{align}
where
${\mathcal B}\big(K_L\to\pi^0\pi^0\nu_\tau\nu_\mu\big)$ is due to both
\,$K_L\to\pi^0\pi^0\nu_\mu\bar\nu_\tau$\, and \,$K_L\to\pi^0\pi^0\nu_\tau\bar\nu_\mu$\,
which have the same rate, as the second line of Eq.\,(\ref{BK2pipi'ff'}) indicates.

Instead of choosing \,$\texttt C_{\nu_\tau\nu_\mu}^{\texttt V,\texttt A}=0$\, which led
to Eq.\,(\ref{vll'}), we also alternatively try to maximize the hyperon rates by allowing
the two different terms in $\texttt C_{\nu_\tau\nu_\mu}^{\texttt V}$ and
$\tilde{\textsf c}_{\nu_\tau\nu_\mu}^{\textsc v}$ as defined in Eq.\,\,(\ref{Cff'a}) to have any
values compatible with the conditions in Eqs.\,\,(\ref{K2pinv}) and\,\,(\ref{yylimit}),
but again with the matrices in Eq.\,(\ref{Y1Y2}).
The branching fractions we find are just marginally larger than those in Eqs.\,\,(\ref{4body})
and\,\,(\ref{B->B'ff'}) and correspond to
\,$|\texttt C_{\nu_\tau\nu_\mu}^{\texttt V}|\simeq1.0\times10^{-10}\rm\,GeV^{-2}$\, and
\,$|\tilde{\textsf c}_{\nu_\tau\nu_\mu}^{\textsc v}|\simeq9.1\times10^{-9}\rm\,GeV^{-2}$.\,

The numbers in Eq.\,(\ref{4body}) are substantially bigger, by a few orders of magnitude, than
their SM counterparts, \,${\mathcal B}(K^-\to\pi^0\pi^-\nu\bar\nu)_{\textsc{sm}}^{}\sim10^{-14}$\,
and \,${\mathcal B}(K_L\to\pi^0\pi^0\nu\bar\nu)_{\textsc{sm}}^{}\sim10^{-13}$
\cite{Kamenik:2011vy,Littenberg:1995zy,Chiang:2000bg}, but still lie significantly under
the existing measured bounds
\,${\mathcal B}(K^-\to\pi^0\pi^-\nu\bar\nu)_{\rm exp}^{}<4.3\times10^{-5}$~\cite{Adler:2000ic}
and ${\mathcal B}(K_L\to\pi^0\pi^0\nu\bar\nu)_{\rm exp}^{}<8.1\times10^{-7}$ \cite{E391a:2011aa}
both at 90\% CL.
Likewise, the hyperon results in Eq.\,(\ref{B->B'ff'}) greatly exceed their SM expectations,
which fall within the $10^{-13}$-$10^{-11}$ range~\cite{Tandean:2019tkm}, but the former are
not yet close to the sensitivity levels of BESIII estimated in Ref.\,\cite{Li:2016tlt} for
the branching fractions of
\,$\Lambda\to n\nu\bar\nu$,\, $\Sigma^+\to p\nu\bar\nu$,\, $\Xi^0\to\Lambda\nu\bar\nu$,\,
$\Xi^0\to\Sigma^0\nu\bar\nu$,\, and \,$\Omega^-\to\Xi^-\nu\bar\nu$,\, which are
\,$3\times10^{-7}$,\, $4\times10^{-7}$,\, $8\times10^{-7}$,\, $9\times10^{-7}$,\, and
\,$2.6\times10^{-5}$,\, respectively.

One could attempt to get around instead the $\tau$ decay constraints by arranging the nonvanishing
elements of the Yukawa matrices to be on, say, the third columns.
It follows that the kaon-mixing restrictions apply to these elements.
This option turns out to be more restraining than Eq.\,(\ref{yylimit}) and hence leads to
comparatively smaller kaon and hyperon rates.
Another possibility would be to enlist the third scalar LQ referred to earlier,
$S_3\, \big(\bar 3,3,1/3\big)$, but we have found that it would not improve on the situation
described in the last paragraph.

The above exercise then suggests that with only SM fermions participating in the interactions of
the scalar LQs it would be unlikely for the \,$|\Delta S|=1$\, hyperon decays
with missing energy to have rates sizable enough to be within the reach of BESIII.
Therefore, future facilities such as super charm-tau factories may be needed to test this
particular LQ scenario.

Here we discuss briefly what we might have to do differently in order to repeat the preceding
steps and simultaneously explain, at least partially, the $B$ anomalies.
It turns out that, among various options, changing only the first of the matrices in
Eq.\,(\ref{Y1Y2}) into
\begin{align} \label{Y1new}
\texttt Y_1^{\textsc{ll}} & \,= \left(\begin{array}{ccc} 0 & 0 & y_{1,13}^{}
\smallskip \\ 0 & y_{1,22}^{} & y_{1,23}^{} \smallskip \\
0 & y_{1,32}^{} & y_{1,33}^{} \end{array}\right) ~~~ ~~~~
\end{align}
may suffice to account for the recent anomalous measurements of \,$b\to s\mu^+\mu^-$\,
transitions within their 2$\sigma$ ranges~\cite{Bhattacharya:2016mcc,Crivellin:2017zlb,
Buttazzo:2017ixm,Sahoo:2015pzk,Sahoo:2018ffv,Kumar:2016omp,Dorsner:2016wpm,Cai:2017wry,
Fajfer:2018bfj,Fayyazuddin:2018zww}.
To exhibit one specific sample set of parameters which can achieve this, we have
\,$\big(y_{1,22}^{},y_{1,23}^{},y_{1,32}^{},y_{1,33}^{}\big)\simeq(-0.2,-0.1,3.3,-0.05)$\, and
\,$m_{S_1}\simeq3.3$\,\,TeV\, from Ref.\,\cite{Cai:2017wry}, along with $y_{1,13}^{}$ values
which comply with Eq.\,(\ref{y1y1limit}) and perturbativity.
For $S_1$ to be responsible also for the \,$b\to c$\, anomalies would however require
the inclusion of its couplings to right-handed SM fermions~\cite{Cai:2017wry}.

\subsection{Including right-handed neutrinos\label{3N}}

In the presence of $\texttt{\textsl N}_y$, we turn on the $\tilde{\textsc y}_1^{\textsc{rr}}$
and $\tilde{\textsc y}_2^{\textsc{lr}}$ terms specified in Eq.\,(\ref{Llqs}).
To fulfill Eq.\,(\ref{K2pinv}) once more, we can first set
\,$\texttt C_{\texttt{\textsl N}\texttt{\textsl N}'}^{\texttt V,\texttt A}=0$\,
in Eq.\,(\ref{Cff'b}), implying that
\begin{align} \label{tcNN'}
\tilde{\textsf c}_{\texttt{\textsl N}\texttt{\textsl N}'}^{\textsc v} & \,=\,
\tilde{\textsf c}_{\texttt{\textsl N}\texttt{\textsl N}'}^{\textsc a}   \,=\,
\frac{-\tilde{\textsc y}_{1,1x}^{\textsc{rr}*}
\tilde{\textsc y}_{1,2y}^{\textsc{rr}}}{4 m_{S_1}^2}
\,=\, \frac{-\tilde{\textsc y}_{2,2x}^{\textsc{lr}*}
\tilde{\textsc y}_{2,1y}^{\textsc{lr}}}{4 m_{\tilde R_2}^2} \,,
\end{align}
but subsequently we shall also touch on the
\,$\texttt C_{\texttt{\textsl N}\texttt{\textsl N}'}^{\texttt V,\texttt A}\neq0$\, case.
As we will see shortly, $\tilde{\textsc y}_1^{\textsc{rr}}$ and
$\tilde{\textsc y}_2^{\textsc{lr}}$ can bring about much larger effects than
$\texttt Y_1^{\textsc{ll}}$ and $\texttt Y_2^{\textsc{rl}}$ discussed in the previous example,
and so we can ignore the contributions of the latter to the \,$s\to d\slashed E$\,
reactions.

Similarly to the first case (A), the Yukawa couplings in Eq.\,(\ref{tcNN'}) affect
$K^0$-$\bar K^0$ mixing through box diagrams, with the LQs and RH neutrinos going around
the loops, giving rise to the effective Hamiltonian
\begin{align}
\tilde{\cal H}_{|\Delta S|=2}^{\textsc{lq}} & \,=\,
\frac{\big( \raisebox{0pt}{\small$\sum_x$} \tilde{\textsc y}_{2,2x}^{\textsc{lr}}
\tilde{\textsc y}_{2,1x}^{\textsc{lr}*} \big)\raisebox{1pt}{$\!^2$}}{128\pi^2\,m_{\tilde R_2}^2}\,
\bar s_L^{}\gamma^\eta d_L^{}\, \bar s_L^{}\gamma_\eta^{}d_L^{}
+ \frac{\big( \raisebox{0pt}{\small$\sum_x$} \tilde{\textsc y}_{1,1x}^{\textsc{rr}}
\tilde{\textsc y}_{1,2x}^{\textsc{rr}*} \big)\raisebox{1pt}{$\!^2$}}{128\pi^2\,m_{S_1}^2}\,
\bar s_R^{}\gamma^\eta d_R^{}\, \bar s_R^{}\gamma_\eta^{}d_R^{}
\,+\, {\rm H.c.}
\end{align}
Evidently, the limitations from kaon-mixing data can again be eluded by assigning the nonzero
elements of the first and second rows of the $\tilde{\textsc y}$ matrices to separate columns.
As for processes which do not conserve charged-lepton flavor, their data no longer offer
pertinent constraints on the $\tilde{\textsc y}$ matrix elements because they are not
associated with SM leptons.

The main restrictions would then come from the empirical bounds on \,$K\to\pi\pi'\slashed E$.\,
To illustrate the implication, we consider this sample set of matrices that satisfy
the above requisites:
\begin{align} \label{ty1ty2} &&
\tilde{\textsc y}_1^{\textsc{rr}} & \,= \left(\begin{array}{ccc} 0 & \tilde y_{1,12}^{} & 0
\smallskip \\ 0 & 0 & \tilde y_{1,23}^{} \smallskip \\ 0 & 0 & 0 \end{array}\right) , &
\tilde{\textsc y}_2^{\textsc{lr}} \,= \left(\begin{array}{ccc} 0 & 0 & \tilde y_{2,13}^{}
\smallskip \\ 0 & \tilde y_{2,22}^{} & 0 \smallskip \\ 0 & 0 & 0 \end{array}\right) . && ~~~
\end{align}
These yield
\begin{align} \label{tcN2N3}
\tilde{\textsf c}_{\texttt{\textsl N}_2\texttt{\textsl N}_3}^{\textsc v} & \,=\,
\tilde{\textsf c}_{\texttt{\textsl N}_2\texttt{\textsl N}_3}^{\textsc a}   \,=\,
\frac{-\tilde y_{1,12}^*\, \tilde y_{1,23}^{}}{4 m_{S_1}^2}
\,=\, \frac{-\tilde y_{2,22}^*\, \tilde y_{2,13}^{}}{4 m_{\tilde R_2}^2} \,,
\end{align}
the other $\tilde{\textsf c}_{\texttt{\textsl f}\texttt{\textsl f}'}^{\textsc v,\textsc a}$
vanishing.\footnote{As before, there are alternatives, such as those with the corresponding
products of nonzero matrix elements being given by
$\big(\tilde y_{1,13}^*\,\tilde y_{1,22}^{},\, \tilde y_{2,23}^*\,\tilde y_{2,12}^{}\big)$,
$\big(\tilde y_{1,11}^*\,\tilde y_{1,23}^{},\, \tilde y_{2,21}^*\,\tilde y_{2,13}^{}\big)$, and
$\big(\tilde y_{1,13}^*\,\tilde y_{1,21}^{},\, \tilde y_{2,23}^*\,\tilde y_{2,11}^{}\big)$.}
Given that the ${\mathcal B}\big(K_L\to\pi^0\pi^0\nu\bar\nu\big){}_{\rm exp}^{}$ bound cited in
the last subsection is stronger than
the ${\mathcal B}(K^-\to\pi^0\pi^-\nu\bar\nu)_{\rm exp}$ one, we can impose
\,${\mathcal B}\big(K_L\to\pi^0\pi^0\slashed E\big){}_{\rm LQ}^{}<8\times10^{-7}$,\,
where
\,${\mathcal B}\big(K_L\to\pi^0\pi^0\slashed E\big){}_{\rm LQ}^{}=
{\mathcal B}\big(K_L\to\pi^0\pi^0\texttt{\textsl N}_2\bar{\texttt{\textsl N}}_3\big) +
{\mathcal B}\big(K_L\to\pi^0\pi^0\texttt{\textsl N}_3\bar{\texttt{\textsl N}}_2\big)
= 2 {\mathcal B}\big(K_L\to\pi^0\pi^0\texttt{\textsl N}_2\bar{\texttt{\textsl N}}_3\big)$.\,
This translates into
\begin{align} \label{K2ppffbound}
\big|\tilde{\textsf c}_{\texttt{\textsl N}_2\texttt{\textsl N}_3}^{\textsc v}\big|
\raisebox{1pt}{$^2$} & \,<\, 9.4\times10^{-14}{\rm\;GeV}^{-4} ~~
\end{align}
and consequently
\begin{align} \label{tytylimit}
\frac{\big|\tilde y_{1,12}^*\, \tilde y_{1,23}^{}\big|}{m_{S_1}^2} & \,=\,
\frac{\big|\tilde y_{2,22}^*\, \tilde y_{2,13}^{}\big|}{m_{\tilde R_2}^2} \,<\,
\frac{1.2}{\rm TeV^2} \,.
\end{align}

As mentioned earlier, the latest search by CMS~\cite{Sirunyan:2018kzh} for scalar LQs
decaying fully into a quark and neutrino has ruled out masses up to 1.1 TeV at 95\%\,\,CL.
This is applicable to the case in which the neutrino is a right-handed one,
$\texttt{\textsl N}{}_y$, and can accommodate Eq.\,(\ref{tytylimit}) because
$|\tilde y_{1,12\,}^{}\tilde y_{1,23}^{}|$ and $|\tilde y_{2,22\,}^{}\tilde y_{2,13}^{}|$
can each have a size up to the perturbativity limit of $4\pi$.
These parameters also enter loop diagrams containing the LQs and SM quarks and contributing to
the invisible partial width of the $Z$ boson, but we have checked that the impact is negligible.

Incorporating Eq.\,(\ref{tytylimit}) into the hyperon decay rates from Sec.\,\ref{amps},
we obtain the maximal branching fractions
\begin{align} \label{B2B'NN'}
{\cal B}\big(\Lambda\to n\texttt{\textsl N}_2\bar{\texttt{\textsl N}}_3\big) & \,<\,
1.3\times10^{-5} \,, &
{\cal B}\big(\Sigma^+\to p\texttt{\textsl N}_2\bar{\texttt{\textsl N}}_3\big) & \,<\,
3.5\times10^{-6} \,, &
\nonumber \\
{\cal B}\big(\Xi^0\to\Lambda\texttt{\textsl N}_2\bar{\texttt{\textsl N}}_3\big) & \,<\,
1.9\times10^{-6} \,, &
{\cal B}\big(\Xi^0\to\Sigma^0\texttt{\textsl N}_2\bar{\texttt{\textsl N}}_3\big) & \,<\,
2.6\times10^{-6} \,, &
\nonumber \\
{\cal B}\big(\Xi^-\to\Sigma^-\texttt{\textsl N}_2\bar{\texttt{\textsl N}}_3\big) & \,<\,
3.2\times10^{-6} \,, &
{\cal B}\big(\Omega^-\to\Xi^-\texttt{\textsl N}_2\bar{\texttt{\textsl N}}_3\big) & \,<\,
1.5\times10^{-4} \,. &
\end{align}
The higher ends of these predictions well exceed their counterparts in Eq.\,(\ref{B->B'ff'}) and
also the corresponding BESIII sensitivity levels~\cite{Li:2016tlt}:
${\cal B}(\Lambda\to n\nu\bar\nu)<3\times10^{-7}$,\,
${\cal B}(\Sigma^+\to p\nu\bar\nu)<4\times10^{-7}$,\,
${\cal B}(\Xi^0\to\Lambda\nu\bar\nu)<8\times10^{-7}$,\,
${\cal B}(\Xi^0\to\Sigma^0\nu\bar\nu)<9\times10^{-7}$,\, and
\,${\cal B}(\Omega^-\to\Xi^-\nu\bar\nu)<2.6\times10^{-5}$.\,
It follows that BESIII might soon uncover NP clues in these processes or, if not, come up with
improved restrictions on the interactions of the scalar LQs with the RH neutrinos.
We remark that the numbers in Eq.\,(\ref{B2B'NN'}) are consistent with the indirect upper bounds
on the branching fractions of yet-unobserved decay modes of $\Lambda$, $\Sigma^+$, $\Xi^0$,
$\Xi^-$, and $\Omega^-$, namely \,$1.4\times10^{-2}$,\, $8.0\times10^{-3}$,\, $3.4\times10^{-4}$,\,
$8.3\times10^{-4}$,\, and \,$1.6\times10^{-2}$,\, respectively,\, which were inferred in
Ref.\,\cite{Su:2019ipw} from the data (at 2 sigmas) on the observed
channels~\cite{Tanabashi:2018oca}.

Instead of fixing
\,$\texttt C_{\texttt{\textsl N}_2\bar{\texttt{\textsl N}}_3}^{\texttt V,\texttt A}=0$\,
which led to Eq.\,(\ref{tcN2N3}), we also try alternatively to maximize the hyperon rates by
allowing the two terms in $\texttt C_{\texttt{\textsl N}_2\bar{\texttt{\textsl N}}_3}^{\texttt V}$
and $\tilde{\textsf c}_{\texttt{\textsl N}_2\bar{\texttt{\textsl N}}_3}^{\textsc v}$ as defined
in Eq.\,(\ref{Cff'b}) to vary freely and at the same time applying the conditions in
Eqs.\,\,(\ref{K2pinv}) and\,\,(\ref{K2ppffbound}),
but still employing the matrices in Eq.\,(\ref{ty1ty2}).
In this instance, the results we get turn out to be barely bigger than those
in Eq.\,(\ref{B2B'NN'}), with
\,$|\texttt C_{\texttt{\textsl N}_2\bar{\texttt{\textsl N}}_3}^{\texttt V}| \simeq
1.0\times10^{-10}\rm\,GeV^{-2}$\,
and
\,$|\tilde{\textsf c}_{\texttt{\textsl N}_2\bar{\texttt{\textsl N}}_3}^{\textsc v}| \simeq
3.1\times10^{-7}\rm\,GeV^{-2}$.\,

The examples presented in this subsection and the previous one indicate that experiments
on these hyperon transitions can serve as a valuable tool to discriminate NP models.
Furthermore, the acquired data could supply information on potential NP contributions which
is complementary to that gained from the kaon sector, including restraints on them which may
be stricter than those provided by kaon measurements.
Lastly, another lesson is that, as the Yukawa matrices in Eq.\,(\ref{ty1ty2}) is generally
independent from that in Eq.\,(\ref{Y1new}), it might happen that distinct segments of the same
LQ scenario could lead to noticeable NP signs in both hyperon and $b$-hadron data.

\section{Conclusions\label{concl}}

We have explored the effects of scalar leptoquarks on $ds\texttt{\textsl f}\texttt{\textsl f}'$
interactions that involve light invisible spin-1/2 fermions, $\texttt{\textsl f}$ and
$\texttt{\textsl f}'$, and induce the FCNC decays of strange hadrons with missing energy,
$\slashed E$, carried away by the fermions.
We concentrate on the case in which the LQ influence may be insignificant on the kaon mode
\,$K\to\pi\slashed E$\, but can be substantial on the hyperon decays
\,$\mathfrak B\to\mathfrak B'\slashed E$\, and \,$\Omega^-\to\Xi^-\slashed E$,\,
as well as on \,$K\to\pi\pi'\slashed E$.\,
This can occur because \,$K\to\pi\slashed E$\, are sensitive exclusively to
$ds\texttt{\textsl f}\texttt{\textsl f}'$ operators with parity-even quark bilinears,
whereas the hyperon modes are affected not only by this kind of operators but also by those
with solely parity-odd quark bilinears.
We show that this possibility can be realized in a model containing scalar LQs by using two
of them which have different chiral couplings to the $d$ and $s$ quarks.
If SM fermions alone participate in the interactions of the LQs, we find that due to
the combined restrictions from kaon mixing and lepton-flavor-violating processes
the hyperon rates cannot reach values likely to be detectable in the near future.
In contrast, with the inclusion of light SM-gauge-singlet right-handed neutrinos which also
couple directly to the LQs, extra \,$s\to d\slashed E$\, channels can arise, with the RH
neutrinos being emitted invisibly, and moreover the constraints from lepton-flavor-violation
data can be evaded.
As a consequence, the resulting hyperon rates are permitted to increase to levels which may be
discoverable by the ongoing BESIII or future efforts such as at super charm-tau
factories.\footnote{It is worth noting that considerable branching fractions of the hyperon modes
\,$\mathfrak B\to\mathfrak B'\slashed E$\, and \,$\Omega^-\to\Xi^-\slashed E$\, are also attainable
if the missing energy is carried away by a single massless dark photon \cite{Su:2019ipw}.}
This in addition suggests that the hyperon sector could serve as another environment in which to
seek sterile neutrinos.
To conclude, our analysis based on simple NP scenarios illustrates the importance of experimental
quests for these hyperon modes, as the outcomes would complement those of measurements on
their kaon counterparts and could help distinguish NP models.

\acknowledgments

This research was supported in part by the MOST (Grant No. MOST 106-2112-M-002-003-MY3).


\begin{thebibliography}{0}

\bibitem{Buchalla:1995vs}
G.~Buchalla, A.J.~Buras, and M.E.~Lautenbacher,
``Weak Decays Beyond Leading Logarithms'',
  Rev.\ Mod.\ Phys.\  {\bf 68}, 1125 (1996) %  doi:10.1103/RevModPhys.68.1125
  [hep-ph/9512380].
  %%CITATION = doi:10.1103/RevModPhys.68.1125;%%

%\cite{Artamonov:2008qb}
\bibitem{Artamonov:2008qb}
  A.V.~Artamonov {\it et al.} [E949 Collaboration],
``New measurement of the $K^{+} \to \pi^{+} \nu \bar{\nu}$ branching ratio'',
  Phys.\ Rev.\ Lett.\  {\bf 101}, 191802 (2008) %  doi:10.1103/PhysRevLett.101.191802
  [arXiv:0808.2459 [hep-ex]].
  %%CITATION = doi:10.1103/PhysRevLett.101.191802;%%

\bibitem{Ahn:2018mvc}
  J.K.~Ahn {\it et al.} [KOTO Collaboration],
``Search for the $K_L\to\pi^0\nu\overline{\nu}$ and $K_L\to\pi^0X^0$ decays at the J-PARC KOTO
experiment'',
  Phys.\ Rev.\ Lett.\  {\bf 122}, no. 2, 021802 (2019) %  doi:10.1103/PhysRevLett.122.021802
  [arXiv:1810.09655 [hep-ex]].
  %%CITATION = doi:10.1103/PhysRevLett.122.021802;%%

\bibitem{CortinaGil:2018fkc}
  E.~Cortina Gil {\it et al.} [NA62 Collaboration],
``First search for $K^+\rightarrow\pi^+\nu\bar\nu$ using the decay-in-flight technique'',
  Phys.\ Lett.\ B {\bf 791}, 156 (2019) %  doi:10.1016/j.physletb.2019.01.067
  [arXiv:1811.08508 [hep-ex]].
  %%CITATION = doi:10.1016/j.physletb.2019.01.067;%%

\bibitem{Tanabashi:2018oca}
  M.~Tanabashi {\it et al.} [Particle Data Group],
``Review of Particle Physics'',
  Phys.\ Rev.\ D {\bf 98}, no. 3, 030001 (2018). % doi:10.1103/PhysRevD.98.030001
  %%CITATION = doi:10.1103/PhysRevD.98.030001;%%

%\cite{Tandean:2019tkm}
\bibitem{Tandean:2019tkm}
  J.~Tandean,
``Rare hyperon decays with missing energy'',
  JHEP {\bf 1904}, 104 (2019) %  doi:10.1007/JHEP04(2019)104
  [arXiv:1901.10447 [hep-ph]].
  %%CITATION = doi:10.1007/JHEP04(2019)104;%%

%\cite{Li:2019cbk}
\bibitem{Li:2019cbk}
  G.~Li, J.Y.~Su, and J.~Tandean,
``Flavor-changing hyperon decays with light invisible bosons'',
  Phys.\ Rev.\ D {\bf 100}, no. 7, 075003 (2019) %  doi:10.1103/PhysRevD.100.075003
  [arXiv:1905.08759 [hep-ph]].
  %%CITATION = doi:10.1103/PhysRevD.100.075003;%%

\bibitem{Adler:2000ic}
  S.~Adler {\it et al.} [E787 Collaboration],
``Search for the decay $K^+\to\pi^+\pi^0\nu\bar\nu$'',
  Phys.\ Rev.\ D {\bf 63}, 032004 (2001) %  doi:10.1103/PhysRevD.63.032004
  [hep-ex/0009055].
  %%CITATION = doi:10.1103/PhysRevD.63.032004;%%

\bibitem{E391a:2011aa}
  R.~Ogata {\it et al.} [E391a Collaboration],
``Study of the $K^0_L\to\pi^0\pi^0\nu\bar\nu$ decay'',
  Phys.\ Rev.\ D {\bf 84}, 052009 (2011) %  doi:10.1103/PhysRevD.84.052009
  [arXiv:1106.3404 [hep-ex]].
  %%CITATION = doi:10.1103/PhysRevD.84.052009;%%

\bibitem{Gninenko:2014sxa}
  S.N.~Gninenko,
``Search for invisible decays of $\pi^0, \eta, \eta', K_S$ and $K_L$:
A probe of new physics and tests using the Bell-Steinberger relation'',
  Phys.\ Rev.\ D {\bf 91}, no. 1, 015004 (2015)  [arXiv:1409.2288 [hep-ph]].
  %%CITATION = doi:10.1103/PhysRevD.91.015004;%%

\bibitem{Li:2016tlt}
  H.B.~Li,
``Prospects for rare and forbidden hyperon decays at BESIII'',
 Front.\ Phys.\ (Beijing) {\bf 12}, no. 5, 121301 (2017) %  doi:10.1007/s11467-017-0691-9
  [arXiv:1612.01775 [hep-ex]]; (Erratum) {\bf 14}, 64001 (2019).
  %%CITATION = doi:10.1007/s11467-017-0691-9;%%

%\cite{Ablikim:2019hff}
\bibitem{Ablikim:2019hff}
M.~Ablikim {\it et al.} [BESIII Collaboration],
``Future Physics Programme of BESIII,''
  Chin.\ Phys.\ C {\bf 44}, no. 4, 040001 (2020) %  doi:10.1088/1674-1137/44/4/040001
  [arXiv:1912.05983 [hep-ex]].
  %%CITATION = doi:10.1088/1674-1137/44/4/040001;%%

%\cite{Bhattacharya:2016mcc}
\bibitem{Bhattacharya:2016mcc}
  B.~Bhattacharya, A.~Datta, J.P.~Gu$\acute{\rm e}$vin, D.~London, and R.~Watanabe,
``Simultaneous Explanation of the $R_K$ and $R_{D^{(*)}}$ Puzzles: a Model Analysis,''
  JHEP {\bf 1701}, 015 (2017) %  doi:10.1007/JHEP01(2017)015
  [arXiv:1609.09078 [hep-ph]].
  %%CITATION = doi:10.1007/JHEP01(2017)015;%%

%\cite{Crivellin:2017zlb}
\bibitem{Crivellin:2017zlb}
  A.~Crivellin, D.~M\"uller, and T.~Ota,
``Simultaneous explanation of R(D$^{(*)}$) and $b\to s\mu^+\mu^-$:
the last scalar leptoquarks standing'',
  JHEP {\bf 1709}, 040 (2017) %  doi:10.1007/JHEP09(2017)040
  [arXiv:1703.09226 [hep-ph]].
  %%CITATION = doi:10.1007/JHEP09(2017)040;%%

%\cite{Buttazzo:2017ixm}
\bibitem{Buttazzo:2017ixm}
  D.~Buttazzo, A.~Greljo, G.~Isidori, and D.~Marzocca,
``B-physics anomalies: a guide to combined explanations'',
  JHEP {\bf 1711}, 044 (2017) %   doi:10.1007/JHEP11(2017)044
  [arXiv:1706.07808 [hep-ph]].
  %%CITATION = doi:10.1007/JHEP11(2017)044;%%

%\cite{Sahoo:2015pzk}
\bibitem{Sahoo:2015pzk}
  S.~Sahoo and R.~Mohanta,
``Lepton flavor violating B meson decays via a scalar leptoquark'',
  Phys.\ Rev.\ D {\bf 93}, no. 11, 114001 (2016) %  doi:10.1103/PhysRevD.93.114001
  [arXiv:1512.04657 [hep-ph]].
  %%CITATION = doi:10.1103/PhysRevD.93.114001;%%

%\cite{Sahoo:2018ffv}
\bibitem{Sahoo:2018ffv}
S.~Sahoo and R.~Mohanta,
``Impact of vector leptoquark on $\bar B \to \bar K^* l^+ l^-$ anomalies'',
  J.\ Phys.\ G {\bf 45}, no. 8, 085003 (2018) %  doi:10.1088/1361-6471/aaca12
  [arXiv:1806.01048 [hep-ph]].
  %%CITATION = doi:10.1088/1361-6471/aaca12;%%

%\cite{Kumar:2016omp}
\bibitem{Kumar:2016omp}
  G.~Kumar,
``Constraints on a scalar leptoquark from the kaon sector'',
  Phys.\ Rev.\ D {\bf 94}, no. 1, 014022 (2016) %  doi:10.1103/PhysRevD.94.014022
  [arXiv:1603.00346 [hep-ph]].
  %%CITATION = doi:10.1103/PhysRevD.94.014022;%%

%\cite{Dorsner:2016wpm}
\bibitem{Dorsner:2016wpm}
  I.~Dor$\check{\rm s}$ner, S.~Fajfer, A.~Greljo, J.F.~Kamenik, and N.~Ko$\check{\rm s}$nik,
``Physics of leptoquarks in precision experiments and at particle colliders'',
  Phys.\ Rept.\  {\bf 641}, 1 (2016) %  doi:10.1016/j.physrep.2016.06.001
  [arXiv:1603.04993 [hep-ph]].
  %%CITATION = doi:10.1016/j.physrep.2016.06.001;%%

%\cite{Cai:2017wry}
\bibitem{Cai:2017wry}
  Y.~Cai, J.~Gargalionis, M.A.~Schmidt, and R.R.~Volkas,
``Reconsidering the One Leptoquark solution: flavor anomalies and neutrino mass'',
  JHEP {\bf 1710}, 047 (2017) %  doi:10.1007/JHEP10(2017)047
  [arXiv:1704.05849 [hep-ph]].
  %%CITATION = doi:10.1007/JHEP10(2017)047;%%

%\cite{Fajfer:2018bfj}
\bibitem{Fajfer:2018bfj}
  S.~Fajfer, N.~Ko$\check{\rm s}$nik, and L.~Vale Silva,
``Footprints of leptoquarks: from $ R_{K^{(*)}} $ to $ K \rightarrow \pi \nu \bar{\nu }$'',
  Eur.\ Phys.\ J.\ C {\bf 78}, no. 4, 275 (2018) %  doi:10.1140/epjc/s10052-018-5757-5
  [arXiv:1802.00786 [hep-ph]].
  %%CITATION = doi:10.1140/epjc/s10052-018-5757-5;%%

%\cite{Fayyazuddin:2018zww}
\bibitem{Fayyazuddin:2018zww}
  Fayyazuddin, M.J.~Aslam, and C.D.~Lu,
``Lepton Flavor Violating Decays of $B$ and $K$ Mesons in Models with Extended Gauge Group'',
  Int.\ J.\ Mod.\ Phys.\ A {\bf 33}, no. 14n15, 1850087 (2018) %  doi:10.1142/S0217751X18500872
  [arXiv:1805.00177 [hep-ph]].
  %%CITATION = doi:10.1142/S0217751X18500872;%%

%\cite{Shanker:1982nd}
\bibitem{Shanker:1982nd}
  O.U.~Shanker,
``$\pi\ell$2, $K\ell$3 and $K^0-\bar K^0$ Constraints on Leptoquarks and Supersymmetric Particles'',
  Nucl.\ Phys.\ B {\bf 204}, 375 (1982). %  doi:10.1016/0550-3213(82)90196-1
  %%CITATION = doi:10.1016/0550-3213(82)90196-1;%%

%\cite{Shanker:1981mj}
\bibitem{Shanker:1981mj}
O.U.~Shanker,
``Flavor Violation, Scalar Particles and Leptoquarks'',
  Nucl.\ Phys.\ B {\bf 206}, 253 (1982). %  doi:10.1016/0550-3213(82)90534-X
  %%CITATION = doi:10.1016/0550-3213(82)90534-X;%%

%\cite{Buchmuller:1986iq}
\bibitem{Buchmuller:1986iq}
  W.~Buchmuller and D.~Wyler,
``Constraints on SU(5) Type Leptoquarks'',
  Phys.\ Lett.\ B {\bf 177}, 377 (1986). %  doi:10.1016/0370-2693(86)90771-9
  %%CITATION = doi:10.1016/0370-2693(86)90771-9;%%

%\cite{Davies:1990sc}
\bibitem{Davies:1990sc}
  A.J.~Davies and X.G.~He,
``Tree Level Scalar Fermion Interactions Consistent With the Symmetries of the Standard Model'',
  Phys.\ Rev.\ D {\bf 43}, 225 (1991). %  doi:10.1103/PhysRevD.43.225
  %%CITATION = doi:10.1103/PhysRevD.43.225;%%

%\cite{Leurer:1993em}
\bibitem{Leurer:1993em}
  M.~Leurer,
``A Comprehensive study of leptoquark bounds'',
  Phys.\ Rev.\ D {\bf 49}, 333 (1994) %  doi:10.1103/PhysRevD.49.333
  [hep-ph/9309266].
  %%CITATION = doi:10.1103/PhysRevD.49.333;%%

%\cite{Leurer:1993qx}
\bibitem{Leurer:1993qx}
 M.~Leurer,
``Bounds on vector leptoquarks'',
  Phys.\ Rev.\ D {\bf 50}, 536 (1994) %  doi:10.1103/PhysRevD.50.536
  [hep-ph/9312341].
  %%CITATION = doi:10.1103/PhysRevD.50.536;%%

%\cite{Davidson:1993qk}
\bibitem{Davidson:1993qk}
  S.~Davidson, D.C.~Bailey, and B.A.~Campbell,
``Model independent constraints on leptoquarks from rare processes'',
  Z.\ Phys.\ C {\bf 61}, 613 (1994) %  doi:10.1007/BF01552629
  [hep-ph/9309310].
  %%CITATION = doi:10.1007/BF01552629;%%

%\cite{Valencia:1994cj}
\bibitem{Valencia:1994cj}
  G.~Valencia and S.~Willenbrock,
``Quark - lepton unification and rare meson decays'',
  Phys.\ Rev.\ D {\bf 50}, 6843 (1994) %  doi:10.1103/PhysRevD.50.6843
  [hep-ph/9409201].
  %%CITATION = doi:10.1103/PhysRevD.50.6843;%%

%\cite{Saha:2010vw}
\bibitem{Saha:2010vw}
  J.P.~Saha, B.~Misra, and A.~Kundu,
``Constraining Scalar Leptoquarks from the K and B Sectors'',
  Phys.\ Rev.\ D {\bf 81}, 095011 (2010) %  doi:10.1103/PhysRevD.81.095011
  [arXiv:1003.1384 [hep-ph]].
  %%CITATION = doi:10.1103/PhysRevD.81.095011;%%

%\cite{Davidson:2010uu}
\bibitem{Davidson:2010uu}
  S.~Davidson and S.~Descotes-Genon,
``Minimal Flavour Violation for Leptoquarks'',
  JHEP {\bf 1011}, 073 (2010) %  doi:10.1007/JHEP11(2010)073
  [arXiv:1009.1998 [hep-ph]].
  %%CITATION = doi:10.1007/JHEP11(2010)073;%%

%\cite{Bobeth:2017ecx}
\bibitem{Bobeth:2017ecx}
  C.~Bobeth and A.J.~Buras,
``Leptoquarks meet $\varepsilon'/\varepsilon$ and rare Kaon processes'',
  JHEP {\bf 1802}, 101 (2018) %  doi:10.1007/JHEP02(2018)101
  [arXiv:1712.01295 [hep-ph]].
  %%CITATION = doi:10.1007/JHEP02(2018)101;%%

%\cite{He:2018uey}
\bibitem{He:2018uey}
  X.G.~He, G.~Valencia, and K.~Wong,
``Constraints on new physics from $K \rightarrow \pi \nu {\bar{\nu }}$'',
  Eur.\ Phys.\ J.\ C {\bf 78}, no. 6, 472 (2018) %  doi:10.1140/epjc/s10052-018-5964-0
  [arXiv:1804.07449 [hep-ph]].
  %%CITATION = doi:10.1140/epjc/s10052-018-5964-0;%%

%\cite{He:2019xxp}
\bibitem{He:2019xxp}
  X.G.~He, J.~Tandean, and G.~Valencia,
``Charged-lepton-flavor violation in $|\Delta S|=1$ hyperon decays'',
  JHEP {\bf 1907}, 022 (2019) %  doi:10.1007/JHEP07(2019)022
  [arXiv:1903.01242 [hep-ph]].
  %%CITATION = doi:10.1007/JHEP07(2019)022;%%

%\cite{Mandal:2019gff}
\bibitem{Mandal:2019gff}
  R.~Mandal and A.~Pich,
``Constraints on scalar leptoquarks from lepton and kaon physics'',
  JHEP {\bf 1912}, 089 (2019) %  doi:10.1007/JHEP12(2019)089
  [arXiv:1908.11155 [hep-ph]].
  %%CITATION = doi:10.1007/JHEP12(2019)089;%%

%\cite{Hirsch:1996qy}
\bibitem{Hirsch:1996qy}
  M.~Hirsch, H.V.~Klapdor-Kleingrothaus, and S.G.~Kovalenko,
``New low-energy leptoquark interactions,''
  Phys.\ Lett.\ B {\bf 378}, 17 (1996) %  doi:10.1016/0370-2693(96)00419-4
  [hep-ph/9602305].
  %%CITATION = doi:10.1016/0370-2693(96)00419-4;%%

%\cite{Dorsner:2017wwn}
\bibitem{Dorsner:2017wwn}
  I.~Doršner, S.~Fajfer, and N.~Ko$\check{\rm s}$nik,
``Leptoquark mechanism of neutrino masses within the grand unification framework,''
  Eur.\ Phys.\ J.\ C {\bf 77}, no. 6, 417 (2017) %  doi:10.1140/epjc/s10052-017-4987-2
  [arXiv:1701.08322 [hep-ph]].
  %%CITATION = doi:10.1140/epjc/s10052-017-4987-2;%%

\bibitem{He:2005we}
  X.G.~He, J.~Tandean, and G.~Valencia,
``Implications of a new particle from the HyperCP data on $\Sigma^+\to p\mu^+\mu^-$'',
  Phys.\ Lett.\ B {\bf 631}, 100 (2005)  [hep-ph/0509041].
  %%CITATION = doi:10.1016/j.physletb.2005.10.005;%%

\bibitem{Bourquin:1981ba}
  M.~Bourquin {\it et al.} [Bristol-Geneva-Heidelberg-Orsay-Rutherford-Strasbourg Collaboration],
``Measurements of Hyperon Semileptonic Decays at the {CERN} Super Proton Synchrotron. 1.
The $\Sigma^- \to \Lambda e^-\bar\nu$ decay mode'',
  Z.\ Phys.\ C {\bf 12}, 307 (1982). %  doi:10.1007/BF01557576
  %%CITATION = doi:10.1007/BF01557576;%%

%\cite{Gaillard:1984ny}
\bibitem{Gaillard:1984ny}
  J.M.~Gaillard and G.~Sauvage,
``Hyperon Beta Decays'',
Ann.\ Rev.\ Nucl.\ Part.\ Sci.\  {\bf 34}, 351 (1984). %  doi:10.1146/annurev.ns.34.120184.002031
  %%CITATION = doi:10.1146/annurev.ns.34.120184.002031;%%

\bibitem{Hsueh:1988ar}
  S.Y.~Hsueh {\it et al.},
``High-precision measurement of polarized-$\Sigma^-$ beta decay'',
  Phys.\ Rev.\ D {\bf 38}, 2056 (1988). %  doi:10.1103/PhysRevD.38.2056
  %%CITATION = doi:10.1103/PhysRevD.38.2056;%%

\bibitem{Dworkin:1990dd}
  J.~Dworkin {\it et al.},
``High statistics measurement of $g_a/g_v$ in $\Lambda\to p + e^- + \bar\nu$'',
  Phys.\ Rev.\ D {\bf 41}, 780 (1990). %  doi:10.1103/PhysRevD.41.780
  %%CITATION = doi:10.1103/PhysRevD.41.780;%%

\bibitem{Batley:2006fc}
  J.R.~Batley {\it et al.} [NA48/I Collaboration],
``Measurement of the branching ratios of the decays $\Xi^0\to\Sigma^+ e^- \bar\nu_e$ and
$\overline{\Xi^0}\to\overline{\Sigma^+} e^+\nu_e$'',
  Phys.\ Lett.\ B {\bf 645}, 36 (2007) %  doi:10.1016/j.physletb.2006.12.028
  [hep-ex/0612043].
  %%CITATION = doi:10.1016/j.physletb.2006.12.028;%%

\bibitem{Hu:2018luj}
  X.H.~Hu and Z.X.~Zhao,
``Study of the $s\to d\nu\bar{\nu}$ rare hyperon decays in the Standard Model and new physics'',
  Chin.\ Phys.\ C {\bf 43}, no. 9, 093104 (2019) %  doi:10.1088/1674-1137/43/9/093104
  [arXiv:1811.01478 [hep-ph]].
  %%CITATION = doi:10.1088/1674-1137/43/9/093104;%%

%\cite{Mahmood:2019bpk}
\bibitem{Mahmood:2019bpk}
  S.~Mahmood,
``Study of Nonstandard Interactions in Rare Decays of Hyprons with Missing Energy'',
  arXiv:1909.13572 [hep-ph].
  %%CITATION = ARXIV:1909.13572;%%

%\cite{He:2019iqf}
\bibitem{He:2019iqf}
  X.G.~He, J.~Tandean, and G.~Valencia,
``Lepton-flavor-violating semileptonic $\tau$ decay and $K\to\pi\nu\bar\nu$'',
  Phys.\ Lett.\ B {\bf 797}, 134842 (2019) %  doi:10.1016/j.physletb.2019.134842
  [arXiv:1904.04043 [hep-ph]].
  %%CITATION = doi:10.1016/j.physletb.2019.134842;%%

%\cite{Aaboud:2019bye}
\bibitem{Aaboud:2019bye}
  M.~Aaboud {\it et al.} [ATLAS Collaboration],
``Searches for third-generation scalar leptoquarks in $\sqrt{s}$ = 13 TeV pp collisions with
the ATLAS detector,''
  JHEP {\bf 1906}, 144 (2019) %  doi:10.1007/JHEP06(2019)144
  [arXiv:1902.08103 [hep-ex]].
  %%CITATION = doi:10.1007/JHEP06(2019)144;%%

%\cite{Aad:2020iuy}
\bibitem{Aad:2020iuy}
  G.~Aad {\it et al.} [ATLAS Collaboration],
``Search for pairs of scalar leptoquarks decaying into quarks and electrons or muons in
$\sqrt{s}=13$ TeV pp collisions with the ATLAS detector,''
  arXiv:2006.05872 [hep-ex].
  %%CITATION = ARXIV:2006.05872;%%

%\cite{Sirunyan:2018kzh}
\bibitem{Sirunyan:2018kzh}
  A.M.~Sirunyan {\it et al.} [CMS Collaboration],
``Constraints on models of scalar and vector leptoquarks decaying to a quark and a neutrino at
$\sqrt{s}=$ 13 TeV,''
  Phys.\ Rev.\ D {\bf 98}, no. 3, 032005 (2018) %  doi:10.1103/PhysRevD.98.032005
  [arXiv:1805.10228 [hep-ex]].
  %%CITATION = doi:10.1103/PhysRevD.98.032005;%%

\bibitem{Littenberg:1995zy}
  L.S.~Littenberg and G.~Valencia,
``The decays $K\to\pi \pi \nu\bar\nu$ within the standard model'',
  Phys.\ Lett.\ B {\bf 385}, 379 (1996) %  doi:10.1016/0370-2693(96)00845-3
  [hep-ph/9512413].
  %%CITATION = doi:10.1016/0370-2693(96)00845-3;%%

\bibitem{Chiang:2000bg}
  C.W.~Chiang and F.J.~Gilman,
``$K_{L,S}\to\pi \pi\nu\bar\nu$ decays within and beyond the standard model'',
  Phys.\ Rev.\ D {\bf 62}, 094026 (2000) %  doi:10.1103/PhysRevD.62.094026
  [hep-ph/0007063].
  %%CITATION = doi:10.1103/PhysRevD.62.094026;%%

\bibitem{Kamenik:2011vy}
  J.F.~Kamenik and C.~Smith,
``FCNC portals to the dark sector'',
  JHEP {\bf 1203}, 090 (2012) %  doi:10.1007/JHEP03(2012)090
  [arXiv:1111.6402 [hep-ph]].
  %%CITATION = doi:10.1007/JHEP03(2012)090;%%

%\cite{Su:2019ipw}
\bibitem{Su:2019ipw}
  J.Y.~Su and J.~Tandean,
``Searching for dark photons in hyperon decays'',
Phys.\ Rev.\ D {\bf 101}, no. 3, 035044 (2020) %  doi:10.1103/PhysRevD.101.035044
  [arXiv:1911.13301 [hep-ph]].
  %%CITATION = doi:10.1103/PhysRevD.101.035044;%%

\end{thebibliography}
\end{document}